\documentclass[10pt]{article}
\pdfoutput=1
\usepackage{jheppub,amsmath,amssymb}
\usepackage{enumerate}
\usepackage{bm}
\usepackage{bbm}
\usepackage{enumitem}
\usepackage[usenames,dvipsnames]{xcolor}
\usepackage{amsmath}
\usepackage{amsfonts}
\usepackage{amssymb}
\usepackage{graphicx}   
\usepackage{hhline}
\usepackage{subfig}
\usepackage{hyperref}
\usepackage{color}
\usepackage{verbatim}
\usepackage{amsmath,amsthm,amssymb}
\usepackage{lscape}
\usepackage{float}
\usepackage{caption}
\usepackage{lscape}
\usepackage{breqn}
\usepackage{multicol}
\usepackage{graphics}
\usepackage{tikz,todonotes}
\usepackage[utf8]{inputenc}
\usepackage{autobreak}
\usepackage{verbatim}
\usepackage{tikz-feynman}
\usepackage{comment}
\usepackage{adjustbox}
\usetikzlibrary{arrows,positioning,decorations.markings,decorations.pathmorphing,calc}
\pgfdeclarelayer{edgelayer}
\pgfdeclarelayer{nodelayer}
\usetikzlibrary{decorations.pathreplacing}
\pgfsetlayers{edgelayer,nodelayer,main}
\tikzset{none/.style={draw=none}}
\tikzset{new edge style 2/.style={black}}
\tikzset{new style 0/.style={black}}
\tikzset{rednode/.style={draw=none, scale=0.3-point,fill=red,circle, draw}}
\tikzset{redline/.style={line width=0.3mm,red}}
\tikzset{greyE/.style={line width=0.1mm,gray}}
\usepackage[utf8]{inputenc}
\usetikzlibrary{arrows,positioning,decorations.markings,decorations.pathmorphing,calc}

\definecolor{hyperref}{RGB}{026,028,087}

\newcommand{\beq}{\begin{equation}}
\newcommand{\eeq}{\end{equation}}
\newcommand{\bea}{\begin{eqnarray}}
\newcommand{\eea}{\end{eqnarray}}
\def\be{\begin{equation}}
\def\ee{\end{equation}}

\def\beq{\begin{equation}}
\def\eeq{\end{equation}}

\def\be{\begin{equation}}
\def\ee{\end{equation}}
\def\ba{\begin{eqnarray}}
\def\ea{\end{eqnarray}}

\def\nn{\nonumber}

\usepackage[normalem]{ulem}

\def\ba{\begin{eqnarray}}
\def\ea{\end{eqnarray}}

\def\D{\mathcal{D}}

\def\({\left(}
\def\){\right)}

\allowdisplaybreaks
\begin{document}

\preprint{Imperial/TP/2021/MC/01}

\title{Massive Double Copy in Three Spacetime Dimensions}

\author[a]{Mariana Carrillo González,}
\author[a]{Arshia Momeni,}
\author[a]{Justinas Rumbutis}

\affiliation[a]{Theoretical Physics, Blackett Laboratory, Imperial College, London, SW7 2AZ, U.K.}

\emailAdd{m.carrillo-gonzalez@imperial.ac.uk}
\emailAdd{arshia.momeni17@imperial.ac.uk}
\emailAdd{j.rumbutis18@imperial.ac.uk}

\abstract{Recent explorations on how to construct a double copy of massive gauge fields have shown that, while any amplitude can be written in a form consistent with colour-kinematics duality, the double copy is generically unphysical. In this paper, we explore a new direction in which we can obtain a sensible double copy of massive gauge fields due to the special kinematics in three-dimensional spacetimes. To avoid the appearance of spurious poles at 5-points, we only require that the scattering amplitudes satisfy one BCJ relation. We show that the amplitudes of Topologically Massive Yang-Mills satisfy this relation and that their double copy at  three, four, and five-points is Topologically Massive Gravity. }

\maketitle


\section{Introduction}
The double copy relation is very powerful since it allows us to obtain gravitational computations from the ``square'' of the simpler analogue Yang-Mills ones. In its better understood form, the double copy consists of a relation between on-shell scattering amplitudes \cite{Kawai:1985xq,Bern:2008qj, Bern:2010ue}, but there have been recent developments showing that this can be further extended to classical results \cite{Saotome:2012vy, Monteiro:2014cda, Luna:2015paa, Luna:2016due, White:2016jzc, Cardoso:2016amd, Luna:2016hge, Goldberger:2017frp, Ridgway:2015fdl, De_Smet_2017, Bahjat_Abbas_2017, Carrillo-Gonzalez:2017iyj,Anastasiou:2014qba,Anastasiou:2016csv,Anastasiou:2017taf,Anastasiou:2017nsz,Anastasiou:2018rdx,Borsten:2015pla,Goldberger_2018, Li_2018, Lee_2018, Plefka_2019, Berman_2019, Kim:2019jwm, Goldberger:2019xef, Alawadhi:2019urr, Banerjee:2019saj, CarrilloGonzalez:2019gof,Shen:2018ebu,Chacon:2021wbr,Bahjat-Abbas:2020cyb,Alfonsi:2020lub,Luna:2020adi,White:2020sfn,Alkac:2021bav,Alawadhi:2020jrv,Huang:2019cja,Keeler:2020rcv,Elor:2020nqe,Farnsworth:2021wvs,Lescano:2021ooe,Ferrero:2020vww,Gumus:2020hbb}. There is an important insight which explains why this relationship holds, the colour-kinematics duality. This duality tells us that the algebra satisfied by the colour part is also satisfied by the kinematic one, thus allowing us to exchange colour for kinematics in order to obtain Gravity from Yang-Mills. The colour-kinematics duality (CK) is explicit in the scattering amplitudes case, but has only been shown to explicitly appear in few classical results. For scattering amplitudes, the double copy at tree level can be proved \cite{Bjerrum-Bohr:2010pnr,Mafra:2011kj,Bjerrum-Bohr:2016axv,delaCruz:2017zqr,Bridges:2019siz}, but the all orders result remains a conjecture. Nevertheless, a large amount of evidence exists that supports this conjecture \cite{Saotome:2012vy,Bern:1997nh,Carrasco:2011mn,Carrasco:2012ca,Mafra:2012kh,Boels:2013bi,Bjerrum-Bohr:2013iza,Bern:2013yya,Bern:2013qca,Nohle:2013bfa,Bern:2013uka,Naculich:2013xa,Mafra:2014gja,Bern:2014sna,Mafra:2015mja,He:2015wgf,Bern:2015ooa,Mogull:2015adi,Chiodaroli:2015rdg,Bern:2017ucb,Oxburgh:2012zr,White:2011yy,Luna:2016idw,SabioVera:2012zky,Johansson:2013nsa,Johansson:2013aca,Bargheer:2012gv,Huang:2012wr,Chen:2013fya,Chiodaroli:2013upa,Johansson:2014zca,Johansson:2017srf,Chiodaroli:2017ehv,Chen:2019ywi,Chi:2021mio,Carrasco:2019yyn,Low:2020ubn,Carrasco:2021ptp} and alternative formulations of the double copy suggest that it also holds at all loop orders \cite{Borsten:2020zgj,Borsten:2021hua}. Most of the examples of a double copy relation involve massless particles, or massless gauge fields plus massive matter \cite{Johansson:2015oia,Bautista:2019tdr,Johansson:2019dnu,Plefka:2019wyg,Bautista:2019evw,Carrasco:2020ywq}. In this paper, we will focus on a less explored case, the double copy of scattering amplitudes involving massive gauge fields.\\

Recent explorations on how to formulate a double copy for massive gauge fields have shown several interesting features. Given that the fields are massive, the generalized gauge transformations allow us to choose kinematic factors that satisfy the same algebra as the colour factors for any theory. Nonetheless, this does not guarantee that the double copy will give rise to a physical theory. While at 4-points this procedure can always be performed \cite{Momeni:2020vvr,Johnson:2020pny}, at 5-points it has been shown that spurious poles can arise \cite{Johnson:2020pny}. A way to avoid these spurious poles was suggested in \cite{Johnson:2020pny}. In this case, instead of considering a single massive gauge boson, one considers a tower of massive fields which satisfy special relations between their masses, these relations were dubbed the spectral conditions. While specific processes in a theory can satisfy these conditions and have a sensible double copy, the only known theories for which any process satisfies the spectral conditions are Kaluza-Klein theories. It has also been shown that if one chooses arbitrary Wilson coefficients for the operators arising from the dimensional reduction, colour-kinematics duality will bring you back to the Kaluza-Klein result \cite{Momeni:2020hmc}. This seems to be too restrictive, but it could be expected since the spectral conditions appear to have been modeled specifically from the knowledge that Kaluza-Klein theories have a sensible double copy given that they satisfy colour-kinematics duality for the massless theories in higher dimensions. In view of this, it is interesting to explore if there are other possibilities in which we can obtain a sensible double copy of massive gauge fields.\\

A question that may arise is if the construction of a massive double copy should follow the massless case as closely as possible. While colour-kinematics should be an ingredient of this procedure, its implications are not completely straightforward. In the massless case, this leads to a specific number of relations between scattering amplitudes called the Bern-Carrasco-Johansson (BCJ) relations. Similarly, in the case of a massive tower of gauge fields satisfying the spectral conditions, one obtains the same number of BCJ relations. Once again, this case models the massless one as closely as possible since it basically arises from the massless case in higher dimensions. We would like to step away from this requirement and consider other options for a sensible massive double copy. More specifically, we would like to understand if the scattering amplitudes of a single gauge field can be double copied. To do so, we will focus on a three-dimensional case, since the special kinematics in 3D have shown to give rise to interesting new examples of the double copy \cite{Huang:2012wr,Bargheer:2012gv}. \\

We will consider topologically massive theories in three spacetime dimensions (3D). These theories consist of the standard Yang-Mills (Einstein-Hilbert) term plus a (gravitational) Chern-Simons term which breaks parity. Both the Topologically Massive Yang-Mills (TMYM) and Topologically Massive Gravity (TMG) have only one degree of freedom. TMG is especially interesting since for a pure Einstein-Hilbert term there are no gravitational degrees of freedom, but when we augment the theory with the gravitational Chern-Simons term, we have one helicity $\pm2$ mode, where the sign depends on the sign of the mass (Chern-Simons) term. It is also worth noting that even though the equations of motion are of third derivative order, the theory is ghost-free and causal. Some explorations of a double copy involving topologically massive gauge theories have been performed in \cite{Moynihan:2020ejh,Burger:2021wss}. By looking at matter particles scattering through a topologically massive mediator in different settings, it has been suggested that a double copy relation exists, but in a non-trivial realization. In this paper, we proceed to analyze the double copy of topologically massive theories without any matter present.\\

The rest of the paper is organized as follows. In Section \ref{sec:spec}, we review the matrix notation for the BCJ double copy, which helps us analyze the appearance of spurious poles. Afterwards, we show how the spurious poles that generically arise in the double copy of a 5-point amplitude can be avoided in three-spacetime dimensions in Section \ref{sec:SpuriousPoles}. We proceed to give an example of theories with a physical double copy up to 5-points; these are the topologically massive theories. In Section \ref{sec:TMYM} and \ref{sec:TMG}, we introduce some relevant aspects of Topologically Massive Yang-Mills and Topologically Massive Gravity respectively. We compute the 3, 4, and 5-point scattering amplitudes and show how TMG corresponds to the double copy of the TMYM. We check the double copy correspondence analytically for the 3 and 4-point amplitudes, and numerically for the 5-point case. We conclude by discussing other interesting double copy relations that arise in 3D and some future directions in Section \ref{sec:discussion}.

\section{Double Copy and BCJ relations}\label{sec:spec}
In this section we review the double copy construction for scattering amplitudes with matrix notation, as introduced in \cite{Momeni:2020hmc}, which is useful to understand the issues that arise when including massive gauge fields. The $n$-point tree level gauge theory amplitude, $A_n$, can be written as
\begin{equation}\label{eq:An}
A_n=g^{n-2}c^T D^{-1}n \ , 
\end{equation}
where $g$ is the coupling strength, $c$ is the vector of colour factors, $D$ is the diagonal matrix with elements given by products of inverse propagators and $n$ is the column vector of kinematic numerators. The Jacobi identities and CK duality in matrix form can be written as
\be
\label{eq:Mc}
Mc=0 \ \rightarrow  \ Mn=0 \ ,
\ee
for a matrix $M$ with entries $\pm1$. Once a representation satisfying CK duality is found, the corresponding amplitude in the gravitational theory, $M_n$, is given as
\begin{equation}\label{eq:Mn}
M_n=i\left(\frac{\kappa}{2}\right)^{n-2}n^T D^{-1} n \ .    
\end{equation}
The kinematic factors directly calculated from Feynman diagrams may not satisfy Jacobi relations, therefore they must be shifted as
\be\label{shift}
n\rightarrow n+\Delta n \ ,
\ee
such that the amplitude in \eqref{eq:An} is unchanged. This can be achieved by setting
\be
\label{eq:deltan}
D^{-1}\Delta n=M^{T}v \ ,
\ee
where $v$ is a vector to be determined. Such shifts are usually referred to as generalized gauge transformations since in the massless case they can be obtained by a gauge transformation and field redefinition of the gauge field. In order to satisfy the CK duality, the shifted $n$ must obey the following equation
\be \label{eq:deltan jac}
M(n+\Delta n)=0\ ,
\ee
which combined with \eqref{eq:deltan} gives
\be\label{eq:cond v}
MDM^{T}v=-Mn\ .
\ee
The number of non-zero rows of $M$ will be equal to the number of Jacobi identities, $N_j$. Therefore, $MDM^{T}$ is block diagonal with a $N_j \times N_j$ symmetric block matrix $A$ and all other elements equal to zero. $Mn$ will have at most $N_j$ non-zero elements so we can write it as
\be \label{eq:U def}
Mn=(U,0, ...,0)\ .
\ee
Note that the vector $U$ measures the violation of the CK algebra. We can see that in order to find the shifts we need to find $v$, for which we need to invert the matrix $A$: 
\be \label{eq:v}
v=-(A^{-1}U,0,..,0)\ .
\ee
It may be that $A$ does not have full rank as it happens in pure Yang-Mills theory. In that case, we can still invert $A$ in the subspace orthogonal to all null eigenvectors of $A$ if $Mn$ is in that subspace, i.e., if there are certain relations between kinematic factors known as BCJ relations.\\

If we now substitute the shifted $n$ back into \eqref{eq:Mn} we get the following:
\begin{equation}
\begin{split}
-i\left(\frac{\kappa}{2}\right)^{-(n-2)}M_n&=(n+\Delta n)^T D^{-1} (n+\Delta n) \\
&=(n+\Delta n)^T D^{-1} n+(n+\Delta n)^T M^T v\\&=n^T D^{-1} n+\Delta n^T D^{-1} n\ ,
\end{split}
\end{equation}
where going from the first to the second line we expanded the expression and used \eqref{eq:deltan}; and going from the second to the third line we used \eqref{eq:deltan jac} to set the last term to zero. Now we can replace $\Delta n$ using \eqref{eq:deltan} and \eqref{eq:v} to get
\begin{equation}\label{eq:Mn final}
\begin{split}
-i\left(\frac{\kappa}{2}\right)^{-(n-2)}M_n=n^T D^{-1} n+v^T M n=n^T D^{-1} n-U^T A^{-1} U \ .   
\end{split}
\end{equation}
Again, in the case when $A$ does not have full rank, $A^{-1}$ and $U$ must be in the subspace orthogonal to the null vectors. Note that the poles of $M_n$ come from kinematic configurations for which either $D$ or $A$ becomes singular. Since $D$ gives rise to the physical poles in the Yang-Mills amplitude, spurious poles could only arise from $A$. \\

As a simple example, lets consider the 4-point amplitude with all massive states with mass $m$ in the adjoint representation. We have a single Jacobi identity, $c_s+c_t+c_u=0$, so $M$ has only one non zero row equal to $(1,1,1)$ and $D=\text{diag}(s-m^2,t-m^2,u-m^2)$. In this case $Mn=(n_s+n_t+n_u,0,0)$, so $U=n_s+n_t+n_u$ and $A$ is $1\times1$ matrix $A=s-m^2+t-m^2+u-m^2=m^2$. Therefore, using \eqref{eq:Mn final} the double copy of the massive 4-point amplitude is
\be \label{eq:4pointDCkin}
-i\left(\frac{\kappa}{2}\right)^{-2}M_4=\frac{n_s^2}{s-m^2}+\frac{n_t^2}{t-m^2}+\frac{n_u^2}{u-m^2}-\frac{(n_s+n_t+n_u)^2}{m^2} \ .
\ee
This shows that no spurious poles arise for the 4-point amplitudes. The 5-point case has been shown to be more complicated and to generically give rise to spurious poles. We analyze the special features that arise for 3D spacetimes in the next section.

\section{Avoiding Spurious Poles in 3D} \label{sec:SpuriousPoles}
In this section, we will analyze the 5-point amplitude for which spurious poles arise generically. We can write the 5-point amplitude as 
\be \label{eq:5point}
A_5=g^3\sum_{i=1}^{15}\frac{c_{i}n_{i}}{s_i-m^2} \ ,
\ee
where the 15 colour factors of the adjoint representation fields are:
\begin{eqnarray}
&& 
c_{1\phantom{0}} \equiv f^{a_1 a_2 b}f^{b a_3 c}f^{c a_4 a_5}\,, \hskip 0.8cm  
c_{2\phantom{1}} \equiv f^{a_2 a_3 b}f^{b a_4 c}f^{c a_5 a_1}\,, \hskip 0.8cm  
c_{3\phantom{1}} \equiv f^{a_3 a_4 b}f^{b a_5 c}f^{c a_1 a_2}\,, \nn \\&&
c_{4\phantom{1}} \equiv f^{a_4 a_5 b}f^{b a_1 c}f^{c a_2 a_3}\,, \hskip 0.8cm  
c_{5\phantom{1}} \equiv f^{a_5 a_1 b}f^{b a_2 c}f^{c a_3 a_4}\,, \hskip 0.8cm  
c_{6\phantom{1}} \equiv f^{a_1 a_4 b}f^{b a_3 c}f^{c a_2 a_5}\,, \nn \\&& 
c_{7\phantom{1}} \equiv f^{a_3 a_2 b}f^{b a_5 c}f^{c a_1 a_4}\,, \hskip 0.8cm  
c_{8\phantom{1}} \equiv f^{a_2 a_5 b}f^{b a_1 c}f^{c a_4 a_3}\,, \hskip 0.8cm  
c_{9\phantom{1}} \equiv f^{a_1 a_3 b}f^{b a_4 c}f^{c a_2 a_5}\,, \nn \\&&
c_{10} \equiv f^{a_4 a_2 b}f^{b a_5 c}f^{c a_1 a_3}\,, \hskip 0.8cm  
c_{11} \equiv f^{a_5 a_1 b}f^{b a_3 c}f^{c a_4 a_2}\,, \hskip 0.8cm  
c_{12} \equiv f^{a_1 a_2 b}f^{b a_4 c}f^{c a_3 a_5}\,, \nn \\\ &&
c_{13} \equiv f^{a_3 a_5 b}f^{b a_1 c}f^{c a_2 a_4}\,, \hskip 0.8cm  
c_{14} \equiv f^{a_1 a_4 b}f^{b a_2 c}f^{c a_3 a_5}\,, \hskip 0.8cm  
c_{15} \equiv f^{a_1 a_3 b}f^{b a_2 c}f^{c a_4 a_5}\,. \hskip 1.5 cm 
\label{FivePointColor}
\end{eqnarray} 
There are 9 independent Jacobi identities that can be written in the form $Mc=0$, where $c=(c_1,..,c_{15})$, and with the matrix $M$ given by
\begin{small}
\begin{eqnarray}
M=\left(
\begin{array}{ccccccccccccccc}
 0 & 0 & 1 & 0 & -1 & 0 & 0 & 1 & 0 & 0 & 0 & 0 & 0 & 0 & 0 \\
 -1 & 0 & 1 & 0 & 0 & 0 & 0 & 0 & 0 & 0 & 0 & 1 & 0 & 0 & 0 \\
 -1 & 0 & 0 & 1 & 0 & 0 & 0 & 0 & 0 & 0 & 0 & 0 & 0 & 0 & 1 \\
 0 & -1 & 0 & 1 & 0 & 0 & 1 & 0 & 0 & 0 & 0 & 0 & 0 & 0 & 0 \\
 0 & -1 & 0 & 0 & 1 & 0 & 0 & 0 & 0 & 0 & 1 & 0 & 0 & 0 & 0 \\
 0 & 0 & 0 & 0 & 0 & -1 & 1 & 0 & 0 & 0 & 0 & 0 & 0 & 1 & 0 \\
 0 & 0 & 0 & 0 & 0 & -1 & 0 & 1 & 1 & 0 & 0 & 0 & 0 & 0 & 0 \\
 0 & 0 & 0 & 0 & 0 & 0 & 0 & 0 & -1 & 1 & 0 & 0 & 0 & 0 & 1 \\
 0 & 0 & 0 & 0 & 0 & 0 & 0 & 0 & 0 & 1 & -1 & 0 & 1 & 0 & 0 \\
 0 & 0 & 0 & 0 & 0 & 0 & 0 & 0 & 0 & 0 & 0 & 0 & 0 & 0 & 0 \\
 0 & 0 & 0 & 0 & 0 & 0 & 0 & 0 & 0 & 0 & 0 & 0 & 0 & 0 & 0 \\
 0 & 0 & 0 & 0 & 0 & 0 & 0 & 0 & 0 & 0 & 0 & 0 & 0 & 0 & 0 \\
 0 & 0 & 0 & 0 & 0 & 0 & 0 & 0 & 0 & 0 & 0 & 0 & 0 & 0 & 0 \\
 0 & 0 & 0 & 0 & 0 & 0 & 0 & 0 & 0 & 0 & 0 & 0 & 0 & 0 & 0 \\
 0 & 0 & 0 & 0 & 0 & 0 & 0 & 0 & 0 & 0 & 0 & 0 & 0 & 0 & 0 \\
\end{array}
\right) \ ,
\end{eqnarray}
\end{small}
Meanwhile, the matrix of propagators, $D$, is given by
\be\label{eq:D}
\begin{split}
 D=&\text{diag}\{ D_{12} D_{45}, D_{15} D_{23}, D_{12} D_{34}, D_{23} D_{45}, D_{15} D_{34}, D_{14} D_{25}, D_{14} D_{23}, D_{25}
D_{34}, \\&D_{13} D_{25}, D_{13} D_{24}, D_{15} D_{24}, D_{12} D_{35}, D_{24} D_{35}, D_{14} D_{35}, D_{13} D_{45}\} \ ,
\end{split}
\ee
where $D_{ij}=-(p_i+p_j)^2-m^2=s_{ij}-m^2$. Using momentum conservation we find that there are only 5 independent Mandelstam invariants; here, we choose them to be $s_{12}$, $s_{13}$, $s_{14}$, $s_{23}$ and $s_{24}$.\\

In order to find the shift of kinematic numerators, $\Delta n$, we need to build and invert the 9x9 matrix  $A$ defined in \eqref{eq:cond v}. In the following, we will keep the discussion general by only using specific features of 3D. Explicit calculation of the determinant of $A$ at 5-points gives:
\begin{equation}
    \text{det}(A)=m^8 (\prod_{i<j} D_{ij}) P(s_{kl},m) \ ,
\end{equation}
where $P(s_{kl},m)$ is a polynomial of the Mandelstam invariants and the mass, given as:
\begin{align} \label{eq:det99}
P(s_{kl},m)=&\,320 m^8+36 m^6 (9 s_{12}+4 (s_{13}+s_{14}+s_{23}+s_{24}))\nonumber\\
&\hspace{-0.5cm}+m^4 \left(117 s_{12}^2+108 s_{12} (s_{13}+s_{14}+s_{23}+s_{24})+4 \left(s_{13} (13 s_{14}+4 s_{23}+17 s_{24})\right.\right.\nonumber\\
&\hspace{1cm}\left.\left.+4
s_{13}^2+4 s_{14}^2+17 s_{14} s_{23}+4 s_{14} s_{24}+4 s_{23}^2+13 s_{23} s_{24}+4
s_{24}^2\right)\right)\nonumber\\
&\hspace{-0.5cm}+2 m^2 \left(9 s_{12}^3+13 s_{12}^2 (s_{13}+s_{14}+s_{23}+s_{24})+s_{12} \left(s_{13} (10 s_{14}+6 s_{23}+17
s_{24})\right.\right.\nonumber\\
&\hspace{1cm}\left.\left.+4 s_{13}^2+4 s_{14}^2+s_{14} (17 s_{23}+6 s_{24})+2 (2 s_{23}+s_{24}) (s_{23}+2 s_{24})\right)\right.\nonumber\\
&\hspace{1cm}\left.+2 \left(s_{13}^2 (s_{14}+2 s_{24})+s_{13}
\left(s_{14}^2+s_{14} (s_{23}+s_{24})+s_{24} (s_{23}+2 s_{24})\right)\right.\right.\nonumber\\
&\hspace{1cm}\left.\left.+s_{23} \left(s_{24} (s_{14}+s_{23})+2 s_{14}
(s_{14}+s_{23})+s_{24}^2\right)\right)\right)\nonumber\\
&\hspace{-0.5cm}+2 s_{24} \left(s_{23} \left(s_{12}^2+s_{12} (s_{13}+s_{14})-s_{13} s_{14}\right)+s_{12}
(s_{12}+s_{13}) (s_{12}+s_{13}+s_{14})\right)\nonumber\\
&\hspace{-0.5cm}+(s_{12} (s_{12}+s_{13}+s_{14})+s_{23} (s_{12}+s_{14}))^2+s_{24}^2
(s_{12}+s_{13})^2 \ ,
\end{align}
with $\prod_{i<j} D_{ij}$ the product of all 10 physical poles. The complicated polynomial, $P(s_{kl},m)$, cannot be expressed as a product of physical poles and it appears in the denominator of $A^{-1}$ of double copy answer in \eqref{eq:Mn final}, therefore it seems that by double copying a generic theory of adjoint fields, all of the same mass gives an unphysical amplitude. However this polynomial has a special structure, it can be expressed as:
\begin{equation}
 P(s_{kl},m)=16\,  \text{det}(p_i\cdot p_j)\ , \quad i,j < 5 \ ,
\end{equation}
where $\text{det}(p_i\cdot p_j)$ is the Gram determinant of the momenta of 4 out of the 5 external states. \\

Note that $P(s_{kl},m)\neq 0$, only if the spacetime dimension is larger than 3 because we cannot have 4 independent vectors in less than four dimensions. Therefore it is zero in our 3D case and $A$ has a null eigenvector, which corresponds to a BCJ relation. First, we need to check if $U$ is orthogonal to that null vector, that is, if the BCJ relation is true. If that is the case, we can invert $A$ in the subspace orthogonal to the null vector. Otherwise, we cannot satisfy colour-kinematics duality. In 3D, the null vector, $e_0$, turns out to have a very simple form:
 \begin{align}
    e_0 &= \begin{bmatrix}
           \epsilon (1,2,3) \\
          -\epsilon (1,2,4)-\epsilon (1,3,4)+\epsilon (2,3,4) \\
           \epsilon (1,2,3)+\epsilon (1,2,4) \\
          -\epsilon (1,3,4)\\
          \epsilon (1,2,4)-\epsilon (2,3,4)\\
           \epsilon (1,2,3)+\epsilon (1,2,4)-\epsilon (2,3,4)\\
           -\epsilon (1,2,4)-\epsilon (1,3,4)\\
           \epsilon (2,3,4)\\
           \epsilon (1,2,3)+\epsilon (1,2,4)+\epsilon (1,3,4) 
         \end{bmatrix}
         \ .
  \end{align}
In the result above, we have expressed the Mandelstam variables in terms of the products of the 3D Levi-Civita tensor and momenta, $\epsilon (i,j,k)=\epsilon_{\mu\nu\sigma}p_i^{\mu}p_j^{\nu}p_k^{\sigma}$, as explained in the Appendix \ref{ap:Relations}. As mentioned before the vector $U$ must satisfy 
\begin{equation}\label{eq:bcj 3D}
    U\cdot e_0=0 \ ,
\end{equation}
in order to be able to satisfy the colour-kinematics duality.\\

For generic kinematics, all 9 components of the null vector are non-zero so we can use the freedom of adding the null eigenvector to $v$ in \eqref{eq:v} to eliminate $v$'s ninth component\footnote{We pick the ninth component as a specific example, but we can alternatively choose any other component since the final answer does not depend on which component we choose.}. 
In other words, we can restrict ourselves to an 8-dimensional subspace to invert the 8x8 submatrix of $A$. This 8x8 matrix, however, still has a complicated polynomial in its determinant:
\begin{equation}
    \text{det}A_{8x8}=-2 m ^6 \left(\prod_{i<j} D_{ij}\right) P_1 (s_{kl},m) \ ,
\end{equation}
with $P_1 (s_{kl},m)$ given by
\begin{equation}
    P_1 (s_{kl},m)=4 (\epsilon (1,2,3)+\epsilon (1,2,4)+\epsilon (1,3,4))^2 \ ,
\end{equation}
where as before we have used the special 3D kinematics from Appendix \ref{ap:Relations} to simplify this expression. At first sight, it appears like we are in trouble again, this polynomial seems to give spurious poles in the double copy of the 5-point amplitude for a generic 3D theory. However we will show that the amplitude does not have spurious poles if \eqref{eq:bcj 3D} is satisfied.\\

One way of seeing the cancellation of the spurious poles is by considering the pseudo-inverse of the matrix $A$:
\be
(A+\varepsilon I)^{-1}=\frac{1}{\text{det}(A+\varepsilon I)}C \ ,
\ee
where $C$ is the cofactor matrix of $(A+\varepsilon I)$. Explicit calculation of these quantities gives the following:
\be
\text{det}(A+\varepsilon I) = -\varepsilon\; 8m^6  \left(\prod_{i<j} D_{ij}\right) e_0 \cdot e_0 + O(\varepsilon^2) \ ,
\ee
and
\be
C_{ij} = -8 m^6 \left(\prod_{i<j} D_{ij}\right)(e_0)_i (e_0)_j+ \varepsilon \bigg(w_i (e_0)_j+w_j (e_0)_i + K_{ij} e_0 \cdot e_0  \bigg)+O(\varepsilon^2) \ ,
\ee
with $w_i$ a vector and $K_{ij}$ a matrix. We can see that in the limit $\varepsilon\rightarrow0$, $U^T (A+\varepsilon I)^{-1} U=\frac{U^T C U}{\text{det}(A+\varepsilon I)}$ is finite if $U$ is orthogonal to $e_0$, that is, if the BCJ relation in \eqref{eq:bcj 3D} is satisfied. Moreover the factor of $e_0\cdot e_0$ in the denominator cancels out since only $K_{ij}$ contributes to the double copy amplitude. This contribution can be expressed as: 
\be \label{eq:UKU}
\begin{split}
\lim_{\varepsilon\rightarrow0} U^T (A+\varepsilon I)^{-1} U=&
-\frac{U^T K U}{8m^6  \left(\prod_{i<j} D_{ij}\right) } \\
=&\sum_{i=1}^{9}\left( \frac{u_i^2}{m^2 D_i}\right)+\frac{(u_1-u_2+u_3-u_4+u_5+u_6-u_7-u_8+u_9)^2}{m^2 D_{35}}\\&+\frac{1}{8 m^2 \epsilon(1,3,4) \epsilon(2,3,4)}\sum_{i=1}^{6}q_i U\cdot e_i
\end{split}
\ee
where $q_i$ is a linear combination of the components of $U$, $e_i$ are 9-component vectors linear in $U$ components and polynomial in Mandelstam variables. Above, we used shorthand notation for the propagators $D_i=\{D_{34},D_{12},D_{45},D_{23},D_{15},D_{14},D_{25},D_{13},D_{24}\}$. \\

Let us first look at the terms in the second line of $\eqref{eq:UKU}$. They contain only physical poles and, as we will see in the following, are consistent with factorization of the scattering amplitude. When $D_i$ goes on shell, \eqref{eq:UKU} goes to $u_i^2/(m^2 D_i)$ where $u_i$ is the violation of the 4pt BCJ relation associated to this factorization channel. The second term in the second line of $\eqref{eq:UKU}$ corresponds to the violation of the 10th Jacobi identity which is not independent from the 9 Jacobi identities encoded in $U=Mn$. Nevertheless, it is needed to have the correct factorization in the $D_{35}\rightarrow0$ limit. As an explicit example, we now illustrate the factorization of the 5pt double copy amplitude, $M_5$, when $D_{12}\rightarrow0$. In this limit the $n^T D^{-1} n$ term in the 5-point double copy \eqref{eq:Mn final} goes to
\be \label{eq:fact nDn}
\frac{1}{D_{12}}\left(\frac{n_1^2}{D_{45}}+\frac{n_3^2}{D_{34}}+\frac{n_{12}^2}{D_{35}}\right),
\ee
which is easy to see since $D$ matrix is diagonal and given in \eqref{eq:D}. Meanwhile, from \eqref{eq:UKU} we find that the $U^T A^{-1} U$ term goes to 
\be \label{eq:fact UKU}
\frac{1}{D_{12}}\left(\frac{u_2^2}{m^2} \right)=\frac{1}{D_{12}}\left(\frac{(-n_1+n_3+n_{12})^2}{m^2} \right),
\ee
since $U$ is a linear combination of the kinematic factors, $U=Mn$ (in particular $u_2=-n_1+n_3+n_{12}$). Using the fact that the 5pt gauge theory amplitude, $A_5$, factorises into the 3-point amplitude $A_3(12I)$ and the 4-point amplitude $A_4(I345)$, where $I$ is an intermediate state., we can write the following expression for the kinematic factors in this limit:
\be \label{eq:fact ns}
n_1=-n_s A_3(12I),\quad n_3=n_t A_3(12I),\quad n_{12}=n_u A_3(12I),
\ee
where $n_s$, $n_t$ and $n_u$ are the kinematic factors of the 4pt amplitude $A_4(I345)$ (if we identify $s_{45}=s$, $s_{34}=t$, $s_{35}=u$ and $c_1=-f^{a_1 a_2 b}c_s$, $c_3=f^{a_1 a_2 b}c_t$, $c_{12}=f^{a_1 a_2 b}c_u$). Substituting \eqref{eq:fact nDn}, \eqref{eq:fact UKU} and \eqref{eq:fact ns} into \eqref{eq:Mn final} we get the following expression for the 5pt double copy amplitude in $D_{12}\rightarrow0$ limit:
\be
M_5\rightarrow
\frac{A_3(12I)^2}{D_{12}}\left(\frac{n_s^2}{s-m^2}+\frac{n_t^2}{t-m^2}+\frac{n_u^2}{u-m^2}-\frac{(n_s+n_t+n_u)^2}{m^2} \right)=\frac{M_3(12I)M_4(I345)}{D_{12}},
\ee
where we used the 3pt double copy relation $M_3(12I)=A_3(12I)^2$.
This shows that the 5pt double copy amplitude correctly factorizes into the 3-point and 4-point double copy amplitudes. This argument can be repeated for all other factorization channels.\\

Now we analyze the third line of \eqref{eq:UKU} which appears to have unphysical poles. However, we will show that the residues of these poles are zero if the BCJ relation in \eqref{eq:bcj 3D} is satisfied. To show this, we note that we can write the third line of \eqref{eq:UKU} with different expressions for $\{q_i,e_i\}$  which correspond to different ways of splitting the answer into $q_i$ and $e_i$. In Appendix \ref{appendix:qe}, we give two explicit expression for $\{q_i,e_i\}$. We have checked numerically that one of these expressions has the property that, for kinematics when $\epsilon(1,3,4)=0$, all $e_i||e_0$. Similarly, for the other expression, all $e_i||e_0$ when $\epsilon(2,3,4)=0$. Therefore, if $U\cdot e_0=0$, then $U\cdot e_i=0$ on the residues of these unphysical poles; so these residues are zero. We conclude that the condition in \eqref{eq:bcj 3D} is sufficient for the double copy to give a physical amplitude at 5-points in 3D with all fields with the same mass and in the adjoint representation. Note that this is quite different from the usual requirement of having 4 BCJ relations like in massless Yang-Mills. Here one relation is enough. Now, an interesting question arises: which theories in three spacetime dimensions satisfy \eqref{eq:bcj 3D}? To tackle this question, in the following sections we will introduce the topologically massive theories. First, we will analyze the 3, 4, and 5-point amplitudes of TMYM and how to write them in terms of kinematic numerators that satisfy the colour-kinematics duality. Afterwards, we will look at the TMG case and show how this corresponds to the double copy of TMYM.

\section{Topologically Massive Yang-Mills} \label{sec:TMYM}
Topologically Massive Yang-Mills theory propagates one spin-1 degree of freedom and is given by a standard Yang-Mills term supplemented with a Chern-Simons term. The TMYM action is \footnote{Note that throughout this paper we use the convention of a mostly plus metric $\eta_{\mu\nu}=(-,+,+)$.}
\be\label{eq:actionTMYM}
S_{TMYM}=\int d^3x\Bigg(-\frac{1}{4}F^{a\mu\nu}F_{a\mu\nu}+\epsilon_{\mu\nu\rho}\frac{m}{12}\left(6 A^{a\mu }\partial^{\nu}A^{\rho}_{a}+g\sqrt{2}f_{abc}A^{a\mu}A^{b\nu}A^{c\rho}\right)\Bigg) \ ,
\ee
where $m$ is the mass of the gauge field and $g$ the coupling strength. Note that the mass term is proportional to the Chern-Simons level. Under gauge transformations, the TMYM is not fully invariant, instead it changes by a total derivative. In this non-Abelian case, the surface integral does not vanish. Instead, it is proportional to the Chern-Simons level times the winding number of the gauge transformation. As it is well known, the invariance of the partition function under gauge transformations leads to the quantization of the Chern-Simons level. In other words, the mass of the gauge field is quantized. We will later see that this feature is not present in the gravitational case.\\

The equations of motion can be easily obtain from \eqref{eq:actionTMYM} and read
\be
D_{\mu} F^{\mu v}+\frac{m}{2} \epsilon^{v \alpha \beta} F_{\alpha \beta}=0 \ ,
\ee
where $D_{\mu}=\partial_{\mu}-\frac{ig}{\sqrt{2}}A_{\mu}$, $F_{\mu\nu}=F_{\mu\nu}^a T^a$,  with $F_{\mu\nu}^a$ the Yang-Mills field strength and $T^a$ the generators of the gauge group. In the following, we choose to work in Lorenz gauge where $\partial_{\mu} A^{\mu}=0$, and $A^{\mu}=A^{\mu \ a}T^a$. It is easy to see that plane waves of the form $A^{\mu }=\varepsilon^\mu e^{ip\cdot x}$ are solutions to the linearised equations of motion as long as the polarisation vectors, $\varepsilon$, satisfy 
\be \label{eom ym}
\varepsilon^{a}_{\mu}+\frac{i}{m}\epsilon_{\mu\nu\rho}p^{\nu}\varepsilon^{a}_{\rho}=0 \ .
\ee
This equation constrains the allowed polarisations of the topologically massive gauge field. Note that we will be denoting the polarisations with $\varepsilon$, to distinguish them from the Levi-Civita symbol denoted with $\epsilon$. Another important ingredient for our scattering amplitudes calculation is the propagator of the gauge field. In an arbitrary gauge the colour stripped propagator is
\be
D_{\mu\nu}[\alpha]=\frac{-i}{p^2+m^2}\left(\eta_{\mu\nu}-\frac{p_{\mu}p_{\nu}}{p^2}-\frac{im}{p^2}\epsilon_{\mu\nu\sigma}p^{\sigma}\right)-\frac{i\alpha}{p^4}p_{\mu}p_{\nu} \ .
\ee
We will work in Landau gauge where $\alpha=0$, hence, 
\be
D_{\mu\nu}=\frac{-i}{p^2+m^2}\left(\eta_{\mu\nu}-\frac{p_{\mu}p_{\nu}}{p^2}-\frac{im}{p^2}\epsilon_{\mu\nu\sigma}p^{\sigma}\right)\ .
\ee
When we look at the gravitational case, we will see how the gravitons propagator can arise as the ``square'' of this one. Given their simplicity, we also present here the three-point and four-point off-shell vertices:
\be
V^{\mu\nu\rho}_{3}=\frac{ig}{\sqrt{2}}f^{a_{1}a_{2}a_{3}}\Bigg(m\epsilon_{\mu\nu\rho}+i\eta^{\mu\nu}(p_{1}^{\rho}-p_{2}^{\rho})+i\eta^{\mu\rho}(p_{3}^{\nu}-p_{1}^{\nu})+i\eta^{\nu\rho}(p_{2}^{\mu}-p_{3}^{\mu})\Bigg) \ ,
\ee
\be
V^{\mu\nu\rho\sigma}_{4}=\frac{ig^2}{2}\Bigg((c_{s}-c_{t})\eta^{\mu\sigma}\eta^{\nu\rho}+(c_{u}-c_{s})\eta^{\mu\rho}\eta^{\nu\sigma}+(c_{t}-c_{u})\eta^{\mu\nu}\eta^{\rho\sigma}\Bigg)\ ,
\ee
where
\be \label{eq:4-pointColor}
c_s=f^{a_1a_2b}f^{ba_3a_4} \ , \quad c_u=f^{a_2a_3b}f^{ba_1a_4} \ , \quad c_t=f^{a_3a_1b}f^{ba_2a_4}  \ .
\ee
In the following we construct the three, four, and five-point amplitudes of TMYM and show how to shift the kinematic numerator so that they satisfy the colour-kinematics duality.

\subsection{TMYM Scattering Amplitudes}
\paragraph{3-point Amplitude}
The three-point on-shell amplitudes is:
\be
A_3=g\Bigg(\sqrt{2}(ee_{13}pe_{12}-ee_{23}pe_{21}+ee_{12}pe_{23})+\frac{im}{\sqrt{2}}\epsilon^{\mu\nu\rho}\varepsilon_{1\mu}\varepsilon_{2\nu}\varepsilon_{3\rho}\Bigg) \ ,
\ee
where we have defined $ee_{ij}\equiv\varepsilon_{i\mu}\varepsilon_{j}^{\mu}$ and $pe_{ij}\equiv p_{i\mu}\varepsilon_{j}^{\mu}$. Using the equation of motion for the polarisation vectors, $\varepsilon_{1}, \varepsilon_{2},\varepsilon_{3}$, as given in \eqref{eom ym}, one can express the first term as follows :
\be\label{eps 3D rel}
(ee_{13}pe_{12}-ee_{23}pe_{21}+ee_{12}pe_{23})=-\frac{3im}{2}\epsilon^{\mu\nu\rho}\varepsilon_{1\mu}\varepsilon_{2\nu}\varepsilon_{3\rho} \ .
\ee
Hence, the 3-point amplitude of topological massive Yang-Mills can be written as:
\be\label{A3}
A_{3}=-ig\sqrt{2}m\epsilon^{\mu\nu\rho}\varepsilon_{1\mu}\varepsilon_{2\nu}\varepsilon_{3\rho} \ .
\ee

\paragraph{4-point Amplitude}
The 4-point TMYM amplitude can be expressed as follows: 
\be
A_4=g^2\left(\frac{c_{s}n_{s}}{s-m^2}+\frac{c_{t}n_{t}}{t-m^2}+\frac{c_{u}n_{u}}{u-m^2}\right)\ ,
\ee
where the colour factor are given in \eqref{eq:4-pointColor} and the kinematic factors are computed using Feynman rules. It is possible to simplify these factors by using the reconstruction methods explained in \cite{ Laurentis:2019bjh, Peraro:2016wsq}. Doing this we find the simpler expressions
\begin{equation}\label{kinematics}
\begin{split}
n_s= \frac{-i}{32s}\Bigg(&ee_{12} ee_{34} (16 m^4 - 12 s^2 + 36 m^2 t - 35 s t - 11 t^2)\\
-& ee_{13} ee_{24} (16 m^4 + 72 m^2 s + 23 s^2 + 11 (4 m^2 + s) t)\\
+&ee_{14} ee_{23} \bigl(-160 m^4 + s (12 s + 11 t) + 4 m^2 (40 s + 11 t)\bigr)\Bigg) \ ,\\
n_t=\frac{i}{8t}\Bigg(&-ee_{12} ee_{34} (4 m^4 + 29 m^2 t + 3 t^2)-ee_{13} ee_{24} (4 m^2-2 s-t) (m^2+3t)\\
+&ee_{14} ee_{23} (4 m^4+29 m^2 t+3 t^2)\Bigg) \ ,\\
n_u= \frac{-i}{32u}\Bigg(&ee_{12} ee_{34} (672 m^4 - 424 m^2 t + (s + t) (12 s + 65 t))\\
-& ee_{13} ee_{24} (672 m^4 - 424 m^2 t -  (41 s - 12 t) (s + t))\\
+&ee_{14} ee_{23}(-848 m^4 + 12 s^2 + 53 s t - 12 t^2 + 8 m^2 (20 s + 33 t))\Bigg)\ .
\end{split}
\end{equation}
These kinematics factors do not satisfy automatically the colour-kinematics duality, i.e. $n_s+n_t+n_u\neq0$. Since we are interested in finding the double copy of TMYM, we need to shift the numerators such that the colour-kinematics duality is satisfied. The new kinematic factors read 
\begin{small}
\be\label{shift n}
\hat{n}_s=n_s-\frac{(n_s+n_t+n_u)(s-m^2)}{(m^2)},\quad \hat{n}_t=n_t-\frac{(n_s+n_t+n_u)(t-m^2)}{(m^2)},\quad
\hat{n}_u=n_u-\frac{(n_s+n_t+n_u)(u-m^2)}{(m^2)} \ ,
\ee
\end{small}
which indeed satisfy the CK duality
\be
\hat{n}_{s}+\hat{n}_{t}+\hat{n}_{u}=\Bigg(\frac{n_s+n_t+n_u}{m^2}\Bigg)\Bigg(4m^2-(s+t+u)\Bigg)=0 \ ,
\ee
given that $s+t+u=4m^2$.

\paragraph{5-point Amplitude}\label{5-point TMYM}
The TMYM 5-point amplitude can be written as in \eqref{eq:5point}. Just as in the previous case, the kinematic factors calculated directly from the Feynman rules do not satisfy the CK algebra. Their explicit expressions are complicated so we do not show them here, but they can be found in the ancillary Mathematica file, FivePointKinematicFactors.m, included in the submission of this paper. The shifted numerators that satisfy the CK duality can be found as explained in Section~\ref{sec:SpuriousPoles}. The 9 component vector $U$ was constructed using \eqref{eq:U def}, and we used the 8x8 submatrix of $A$, $A_{8x8}$ for constructing the shifted kinematic factors. In the construction of $A_{8x8}$, we can eliminate any arbitrary $n$th column and $n$th row since the final result does not depend on this.

\section{Topologically Massive Gravity and the Double Copy} \label{sec:TMG}
The action for Topologically Massive Gravity is 
\be\label{eq:TMG}
S_{TMG}=\frac{1}{\kappa^2}\int d^3x\sqrt{-g}\left(-R-\frac{1}{2m}\epsilon^{\mu\nu\rho}\left(\Gamma^{\alpha}_{\mu\sigma}\partial_{\nu}\Gamma^{\sigma}_{\alpha\rho}+\frac{2}{3}\Gamma^{\alpha}_{\mu\sigma}\Gamma^{\sigma}_{\nu\beta}\Gamma^{\beta}_{\rho\alpha}\right)\right) \ .
\ee
Note that the sign of the Einstein-Hilbert term is the opposite to the conventional one; this is required so that the physical spin-2 mode is not ghostly. The equations of motion are given by
\be
G^{\mu\nu}+\frac{1}{m}C^{\mu\nu}=0 \ ,
\ee
where $G^{\nu\mu} \equiv R^{\nu \mu}-\frac{1}{2} R g^{\nu \mu}$ is the Einstein tensor and $C^{\mu \nu} \equiv \varepsilon^{\mu \alpha \beta} \nabla_{\alpha}\left(R_{\beta}^{\nu}-\frac{1}{4} g_{\beta}^{\nu} R\right)$ the Cotton tensor which is the 3D analogue of the Weyl tensor. Contrary to the TMYM case, in this case the action is fully invariant under diffeomorphisms and no quantization of the mass is required.\\

We now proceed to analyze some elements that are required for the scattering amplitude computations, and how these elements themselves can be constructed as a double copy of the analogue Yang-Mills object. First, we obtain the linearised equations of motion by expanding around flat space as  $g_{\mu\nu}=\eta_{\mu\nu}+\kappa h_{\mu\nu}$. We will work in de Donder gauge where $\partial_{\mu} h^{\mu \nu}-\frac{1}{2} \partial^{\nu} h=0$. As in the Yang-Mills case, the plane wave solution $h_{\mu\nu}=\varepsilon_{\mu\nu}e^{i p \cdot x}$ is a solution of the linearised equations of motion when
\be \label{tmg eom}
\Bigg(\delta^{\mu}_{\alpha}\delta^{\nu}_{\beta}+\frac{i}{2m}(\epsilon^{\nu}_{\rho\sigma}p^{\rho}\delta^{\sigma}_{\alpha}\delta^{\mu}_{\beta}+\epsilon^{\mu}_{\rho\sigma}p^{\rho}\delta^{\sigma}_{\alpha}\delta^{\nu}_{\beta})\Bigg)\varepsilon^{\alpha\beta}=0 \ .
\ee
This again restricts the allowed polarisations of the massive graviton. It is interesting to notice that we can already see a double copy relation at this level. The on-shell polarisation tensors of TMG can be written as the square of the on-shell polarisation vector of TMYM 
\be
\varepsilon_{\mu\nu}=\varepsilon_{\mu}\varepsilon_{\nu} \ .
\ee
If $\varepsilon_{\mu}$ satisfies \eqref{eom ym} then the polarisation tensor defined above will satisfy \eqref{tmg eom}. When we write our scattering amplitudes below, we will be using this relation and writing them in terms of the polarisation vectors $\varepsilon_{\mu}$.\\

It is also instructive to look at the propagator of topologically massive gravitons. In an arbitrary gauge this reads
\be
\D^{\mu\nu\rho\sigma}=D^{\mu\nu\rho\sigma}+\alpha\frac{i}{4q^2}\Bigg(-4\eta^{\mu(\sigma}\eta^{\rho)\nu}+\left(2\eta^{\nu(\sigma}P_{1}^{\rho)\mu}+\nu\sigma\leftrightarrow \mu\rho\right)\Bigg) \ ,
\ee
with
\be
\begin{split}
D^{\mu\nu\rho\sigma}=\frac{i}{p^2+m^2}&\Bigg(-3 \eta^{\mu\nu}\eta^{\rho\sigma}+\Bigg(\eta^{\rho\sigma}P_{1}^{\mu\nu}+\mu\nu\leftrightarrow \rho\sigma\Bigg)+2P_{1}^{\mu(\sigma}P_{1}^{\rho)\nu}-\Bigg(P_{2}^{\mu(\rho}P_{1}^{\sigma)\nu}+P_{2}^{\nu(\rho}P_{1}^{\mu)\sigma}\Bigg)\\
&-\frac{m^2}{p^2}\Bigg(2\eta^{\mu\nu}\eta^{\rho\sigma}-2\eta^{\nu(\sigma}P_{1}^{\rho)\mu}+\frac{2}{p^2}\eta^{\mu(\sigma}p^{\rho)}p^{\nu}\Bigg)-\frac{1}{p^4}p^{\mu}p^{\nu}p^{\rho}p^{\sigma}\Bigg) \ ,
\end{split}
\ee
where $X^{a(b}X^{c)d}=\frac{1}{2}(X^{ab}X^{cd}+X^{ac}X^{bd})$ and
\be
P_{1}^{ab}=\eta^{ab}-\frac{p^{a}p^{b}}{p^2}, \quad P_{2}^{ab}=P_{1}^{ab}-\frac{im}{p^2}\epsilon^{abc}p_{c} \ .
\ee
As before, we will work in de Donder gauge where  $\alpha=0$; hence $\D^{abcd}=D^{abcd}$. A double copy between the propagators of TMYM and TMG has been proposed in \cite{Moynihan:2020ejh}. It was shown that they can be related as
\be
D_{\mu \nu \rho \sigma}=\left(p^{2}+m^{2}\right) D_{\rho(\mu} D_{\nu) \sigma} \ ,
\ee
which matches the TMG case if we ignore terms that vanish when they are contracted to conserved currents. For the gravitational case, we do not show the explicit expression of the off-shell vertices since this is quite involved and gives no insights to our discussion. We will now go ahead and compute the three, four, and five -point amplitudes of TMG and show how they correspond to the double copy of TMYM.

\subsection{TMG Scattering Amplitudes}
\paragraph{3-point Amplitude}
The 3-point amplitude of TMG can be simplified by using the 3-point relations in Appendix \ref{ap:Relations}, which arise from the polarisation vector equations of motion. After using these relations, the 3-point TMG amplitude can be written as follows:
\be
M_3=2i ee_{12} ee_{13} ee_{23}m^2 \kappa \ ,
\ee
which is equivalent to:
\be\label{M3}
M_3=-i\left(\epsilon^{\mu\nu\rho}\epsilon_{1\mu}\epsilon_{2\nu}\epsilon_{3\rho}\right)^2 m^2\kappa\ .
\ee
From this, we can see that the three-point double copy relation
\be
M_3=i\frac{\kappa}{2}A_3 A_3 \ ,
\ee
is satisfied with $A_3$ given by \eqref{A3} and $M_3$ by \eqref{M3}. We conclude that at 3-points, the double copy of TMYM is TMG.

\paragraph{4-point Amplitude}
We compute the 4-point amplitude directly from the Feynman rules and, given its length, only show our result in the Appendix \ref{ap:4pTMG}. Here, we focus on understanding if it corresponds to the TMYM double copy. At 4-points, the double copy relation is the following:
\be\label{4-point dc}
M_4=i\left(\frac{\kappa}{2}\right)^2\Bigg(\frac{\hat{n}^2_s}{s-m^2}+\frac{\hat{n}^2_t}{t-m^2}+\frac{\hat{n}^2_u}{u-m^2}\Bigg) \ ,
\ee
where $\hat{n}$ satisfies the colour-kinematics duality. To find the TMYM double copy, we plug-in the kinematic numerators from \eqref{shift n}. The analytic expressions obtained for the TMYM double copy are highly involved and complicated to simplify. In this case, we can simplify them by using the Breit coordinate system. It is well known that it is advantageous to use  this coordinate system to investigate the analytic properties of the amplitude (see e.g.\cite{Bogolyubov:1959bfo}). For elastic scattering processes, the momenta in the Breit coordinate system are defined as
\begin{equation}\label{breit1}
    \begin{split}
        & p_1^{\mu}=(\sqrt{\vec{p}^{\,2}+m^2}, \vec{p}^{\,}) \ , \quad \ \ p_2^{\mu}=(E, -\vec{p}^{\,}+\lambda\vec{e}^{\,})\ ,\\
        & p_3^{\mu}=(\sqrt{\vec{p}^{\,2}+m^2},-\vec{p}^{\,})\ , \quad p_4^{\mu}=(E, \vec{p}^{\,}+\lambda\vec{e}^{\,})\ ,
    \end{split}
\end{equation}
where $\vec{e}^{\,}\cdot\vec{p}^{\,}=0$, $|\vec{e}^{\,}|=1$, and the arrow denotes 2-dimensional spatial vectors. Note that all momenta are incoming. We choose the following directions, $\vec{p}^{\,}=(p,0)$ and $\vec{e}^{\,}=(0,1)$. In terms of Mandelstam variables we have,
\begin{equation}\label{breit2}
    \begin{split}
    &t=-4p^2,\quad E=\sqrt{p^2+m^2+\lambda^2}=\frac{s-2m^2+t/2}{\sqrt{4m^2-t}},
    \end{split}
\end{equation}
and the external polarisations are obtained from \eqref{pola values}. The explicit expressions for the shifted numerators in this coordinate system are given in \eqref{breit n}. Once we simplify the double copy of TMYM using these relations, we find that it indeed corresponds to the 4-point amplitude in \eqref{4pt Breit}. Therefore, we conclude that at 4-point TMG is the double copy of TMYM. \\

Another way to check the double copy of TMYM that does not require a specific coordinate system is to use the simplified kinematic numerators in \eqref{kinematics}. However, it is still is complicated to see that TMG is the double copy of TMYM. This can only be seen once we relate the product of polarisations and Mandelstam variables as shown in \eqref{4-pointdc}. These relations are satisfied on-shell and were obtained by using random on-shell momenta and polarisation vectors. Our numerical method used to obtain these random kinematics is explained in Appendix \ref{num meth}. Therefore, we can conclude that once \eqref{4-pointdc} is imposed, the 4-pt TMG is the double copy of TMYM.

\paragraph{5-point Amplitude}
At 5-points, we have shown that the double copy in 3D has no spurious poles as long as \eqref{eq:bcj 3D} is satisfied. We have verified numerically that the TMYM vector $U$ satisfies this equation for multiple random on-shell kinematics. Thus, there are no spurious poles in the double copy of the TMYM 5-point amplitude. We obtained the 5-point TMYM double copy using  \eqref{eq:Mn final} and compared it with the 5-point amplitude of TMG computed from Feynman rules. Because of the complexity of the analytic expression of the TMG 5-point amplitude, we only compared the values of both amplitudes evaluated on random kinematic configurations. Some examples of these values are given in Table \ref{tab:valuesMandelstam}. We found that they agree exactly, further confirming the absence of spurious poles in the double copy of the 5-point TMYM amplitude.

\section{Discussion} \label{sec:discussion}
The special kinematics arising in a three-dimensional spacetime allow us to construct a well-defined massive double copy that does not require a tower of massive states. Here, we have shown how the spurious poles that generically appear in 5-point amplitudes can be avoided with a single BCJ relation. This BCJ relation was written in terms of the kinematic numerators, or more precisely, in terms of the breaking of the CK algebra for the kinematic factors, i.e., the vector $U$. It is possible to rewrite this relation in terms of partial amplitudes; nevertheless, the expression for the BCJ relation is largely involved as seen in \eqref{eq:BCJpartialA}. In this paper, we gave an explicit example of a theory that satisfies such BCJ relation and analyzed its double copy. We found that the double copy of Topologically Massive Yang-Mills is well defined and corresponds to Topologically Massive Gravity. The expressions for these scattering amplitudes become quite involved at higher points, and required the use of numerical methods to verify our results. It is quite likely that there are better variables in which the double copy relation becomes cleaner. For example, it would interesting to see if using spinor-helicity variables, similar to \cite{Moynihan:2020ejh}, can lead to more compact expressions that allow to prove the 5-point amplitude double copy analytically.  Throughout this paper, we considered the BCJ version of the double copy, but it would be interesting to see how the KLT analogue is modified. Similarly, one could try to understand if simplifications could arise by writing the results using the formalism of \cite{Chi:2021mio} or \cite{Carrasco:2019yyn,Low:2020ubn,Carrasco:2021ptp}, or if generalizations of our double copy example can be found.  \\

The fact that topologically massive theories satisfy a double copy relation makes us ponder how do these theories fit in the larger web of relations for scattering amplitudes \cite{Cheung:2017ems}. Examples of scalar effective field theories in this web have been shown to satisfy a double copy relation which is inherited from that of YM and gravity. It is interesting to note that a common feature of these theories is that they exhibit conformal invariance at a given spacetime dimension \cite{Cheung:2020qxc,Farnsworth:2021ycg}. In our case, while the gravitational Chern-Simmons term is conformally invariant, the whole action including the Einstein-Hilbert term is not. Similarly, TMYM does not have conformal invariance. It isn't clear if there is a similar feature or an obvious way of relating these theories to the broader web of amplitude relations, but it would be interesting to explore that possibility. Similarly, it is compelling to understand how does the different versions of the classical double copy work for topologically massive theories? There are some simple and some more involved new extensions that we will analyze in future work. It is worth pointing out that in \cite{Burger:2021wss}, perturbative classical results pointed to a not so straightforward double copy realization.\\

Besides TMYM, it is interesting to understand if other 3D theories can also have a well-defined double copy, and if these theories need to satisfy the BCJ relation in \eqref{eq:bcj 3D}. To understand this, we can explore the simple case of massive Yang-Mills in 3D. By following the procedure in Section \ref{sec:SpuriousPoles}, we can analyze the 5-point massive Yang-Mills amplitude. As in 4D, the local numerical factors calculated directly from the Feynman rules do not satisfy the colour-kinematics duality, i.e., \eqref{eq:Mc} is not satisfied. In order to satisfy CK duality, we need to perform the shifts \eqref{shift} and solve \eqref{eq:cond v} to find $v$. However, if we  consider the amplitude with external polarisations that satisfy the TMYM equations of motion, that is, \eqref{eom ym}, the local kinematics of massive Yang-Mills calculated directly from the Feynman rules satisfy the colour-kinematics duality and do not require any shifts. In that case, the double copy will only have physical poles. This shows that the polarisations of TMYM play a special role in giving rise to a physical double copy. It would be interesting to understand if these are the only polarisations that remove the spurious poles in the massive Yang-Mills double copy, or if a generalized procedure should be found in order to construct a double copy for more general polarisations. \\

So far, the massive double copy has only been explored for tree-level processes. It will be intriguing to understand how this generalizes to loop order. A simpler task towards this goal consists of understanding higher corrections in the eikonal limit. At tree-level, we can already see an interesting structure arising. The 4-point amplitudes of the TMYM and TMG in the eikonal limit, $s\rightarrow\infty$ and $t<<m^2$,  read
\be
A^{\scriptscriptstyle TYM}=g^2\frac{\frac{m s}{\sqrt{t}}c_t}{-m^2} \ , \quad
iM^{\scriptscriptstyle TMG}=\frac{-\kappa^2}{2}\frac{s^2}{t}=2\left(\frac{\kappa}{2}\right)^2\frac{\left(\frac{m s}{\sqrt{t}}\right)^2}{-m^2}\ .
\ee
We can immediately observe a double copy relation arising, given that in this limit the propagator is $t-m^2\rightarrow-m^2$. This is not the standard relation, instead it has an extra factor of 2. In fact, it is straightforward to understand this extra factor. In the 4-point double copy, \eqref{eq:4pointDCkin}, the kinematic factor $n_t$ dominates in the eikonal limit which gives rise to $iM=2\left(\frac{\kappa}{2}\right)^2\frac{n_t^2}{m^2}$. Given this new feature arising already at tree-level, we would like to understand how this procedure works at higher orders in the eikonal limit. We leave this analysis for future work.\\

The existence of a 3D massive double copy opens a path for a profuse amount of questions as we have discussed above. We hope that this exploration will stimulate future explorations to advance our understanding of the applicability of the double copy.

\section{Acknowledgments}
We would like to thank Andrew J. Tolley for useful discussions and comments. MCG is supported by the European Union’s Horizon 2020 Research Council grant 724659 MassiveCosmo ERC–2016–COG and the STFC grants ST/P000762/1 and ST/T000791/1. JR is supported by an STFC studentship.

\appendix 
\section{Explicit Expressions of $q_i$ and $e_i$}\label{appendix:qe}
In this appendix we give the explicit expressions for $q_i$ and $e_i$ which were introduced in \eqref{eq:UKU}. Since we can split this expression in different ways, we have the freedom to choose different expressions for $q_i$ and $e_i$. Here we show two different cases that are useful to understand the residues at the spurious poles. One possible choice for $q_i$ and $e_i$ is 
\begin{align*}
q_1&=-u_2-u_4-u_7+u_9 \ ,\\
q_2&=0\ ,\\
q_3&=-u_2-u_4-u_7+u_9\ ,\\
q_4&=u_1-u_2+u_3-u_4+u_6-u_7+2 u_9\ ,\\
q_5&=u_2+u_3+2 u_4+u_5+u_6+u_7-u_9 \ ,\\
q_6&=u_3+u_4+u_5+u_6 \ ,
\end{align*}
\begin{tiny}
 \begin{align*}
    &e_1 =\frac{1}{2} \begin{bmatrix}
           -m^2 \left(15 s_{12}+17 s_{13}+4 s_{14}+10 s_{23}+6 s_{24}\right)+46 m^4+s_{12}^2+2 s_{13}^2+s_{14} s_{23}+s_{13} s_{24}+s_{12} \left(4 s_{13}+s_{14}+s_{23}+s_{24}\right) \\
          -m^2 \left(-7 s_{12}-11 s_{13}-14 s_{14}-7 s_{23}-3 s_{24}\right)-31 m^4-2 s_{13}^2+s_{12} \left(-2 s_{13}-3 s_{14}\right)-4 s_{13} s_{14}-2 s_{13} s_{23}-3 s_{14} s_{23}+s_{13} s_{24}-2 s_{14} s_{24} \\
          -m^2 \left(8 s_{12}+13 s_{13}-5 s_{14}+4 s_{23}+4 s_{24}\right)+32 m^4+2 s_{13}^2-2 s_{14}^2+s_{12} \left(3 s_{13}+s_{14}\right) \\
          0\\
          -m^2 \left(7 s_{12}+11 s_{13}+14 s_{14}+7 s_{23}+3 s_{24}\right)+31 m^4+2 s_{13}^2+4 s_{13} s_{14}+s_{12} \left(2 s_{13}+3 s_{14}\right)+2 s_{13} s_{23}+3 s_{14} s_{23}-s_{13} s_{24}+2 s_{14} s_{24}\\
           -m^2 \left(22 s_{12}+28 s_{13}+18 s_{14}+17 s_{23}+9 s_{24}\right)+77 m^4+s_{12}^2+4 s_{13}^2+4 s_{13} s_{14}+2 s_{13} s_{23}+4 s_{14} s_{23}+2 s_{14} s_{24}+s_{12} \left(6 s_{13}+4 s_{14}+s_{23}+s_{24}\right)\\
           -m^2 \left(7 s_{12}+4 s_{13}+9 s_{14}+6 s_{23}+2 s_{24}\right)+14 m^4+s_{12}^2+2 s_{14}^2+s_{14} s_{23}+s_{13} s_{24}+s_{12} \left(s_{13}+s_{23}+s_{24}\right)\\
           -m^2 \left(-14 s_{12}-15 s_{13}-23 s_{14}-13 s_{23}-5 s_{24}\right)-45 m^4-s_{12}^2-2 s_{13}^2-2 s_{14}^2-4 s_{13} s_{14}-2 s_{13} s_{23}-4 s_{14} s_{23}+s_{12} \left(-3 s_{13}-3 s_{14}-s_{23}-s_{24}\right)-2 s_{14} s_{24}\\
           -m^2 \left(8 s_{12}+13 s_{13}-5 s_{14}+4 s_{23}+4 s_{24}\right)+32 m^4+2 s_{13}^2-2 s_{14}^2+s_{12} \left(3 s_{13}+s_{14}\right) 
         \end{bmatrix} \ ,\\
         &e_2 =\begin{bmatrix}
           m^2 \left(2 s_{12}+2 s_{23}+2 s_{24}\right)-10 m^4+s_{13} s_{14} \\
          m^2 \left(-2 s_{12}-5 s_{14}-2 s_{23}-2 s_{24}\right)+10 m^4+s_{14} \left(s_{12}+s_{13}+s_{23}+s_{24}\right)\\
          m^2 \left(2 s_{12}-4 s_{14}+2 s_{23}+2 s_{24}\right)-10 m^4+s_{14} \left(s_{13}+s_{14}\right) \\
          0\\
          m^2 \left(2 s_{12}+5 s_{14}+2 s_{23}+2 s_{24}\right)-10 m^4+s_{14} \left(-s_{12}-s_{13}-s_{23}-s_{24}\right)\\
           m^2 \left(4 s_{12}+5 s_{14}+4 s_{23}+4 s_{24}\right)-20 m^4+s_{14} \left(-s_{12}-s_{23}-s_{24}\right)\\
           s_{14} \left(4 m^2-s_{14}\right)\\
-m^2 \left(2 s_{12}+9 s_{14}+2 \left(s_{23}+s_{24}\right)\right)+10 m^4+s_{14} \left(s_{12}+s_{13}+s_{14}+s_{23}+s_{24}\right)\\
          2 m^2 \left(s_{12}-2 s_{14}+s_{23}+s_{24}\right)-10 m^4+s_{14} \left(s_{13}+s_{14}\right) 
         \end{bmatrix} \ ,
\\
&e_3 =\frac{1}{2} \begin{bmatrix}

           -m^2 \left(7 s_{12}+5 s_{13}+2 \left(2 s_{14}+s_{23}+s_{24}\right)\right)+10 m^4+s_{12}^2+s_{14} s_{23}+s_{13} \left(2 s_{14}+s_{24}\right)+s_{12} \left(2 s_{13}+s_{14}+s_{23}+s_{24}\right) \\
          -m^2 \left(s_{12}+3 s_{13}+4 s_{14}+s_{23}+s_{24}\right)+5 m^4+s_{12} s_{14}+s_{14} s_{23}+s_{13} \left(2 s_{14}+s_{24}\right)\\
          \left(s_{13}-s_{14}\right) \left(-\left(m^2-s_{12}\right)\right) \\
          0\\
4 m^2 s_{14}+m^2 s_{23}+m^2 s_{24}+s_{12} \left(m^2-s_{14}\right)+s_{13} \left(3 m^2-2 s_{14}-s_{24}\right)-5 m^4-s_{14} s_{23}\\
           \left(m^2-s_{12}\right) \left(5 m^2-s_{12}-2 s_{13}-s_{23}-s_{24}\right)\\
           -m^2 \left(7 s_{12}+4 s_{13}+5 s_{14}+2 s_{23}+2 s_{24}\right)+10 m^4+s_{12}^2+s_{14} s_{23}+s_{13} \left(2 s_{14}+s_{24}\right)+s_{12} \left(s_{13}+2 s_{14}+s_{23}+s_{24}\right)\\
-\left(m^2-s_{12}\right) \left(5 m^2-s_{12}-s_{13}-s_{14}-s_{23}-s_{24}\right)\\
          \left(s_{13}-s_{14}\right) \left(-\left(m^2-s_{12}\right)\right) 
         \end{bmatrix} \  ,
        \\ 
&e_4=\begin{bmatrix}

-2 m^2 \left(s_{12}+s_{13}+s_{23}\right)+8 m^4+s_{13} s_{23}\\

m^2 \left(11 s_{12}+6 s_{13}+4 s_{14}+11 s_{23}+4 s_{24}\right)-28 m^4-s_{13} \left(s_{12}+s_{23}+s_{24}\right)-\left(s_{12}+s_{23}\right) \left(s_{12}+s_{14}+s_{23}+s_{24}\right)\\

-m^2 \left(11 s_{12}+6 s_{13}+2 s_{14}+4 s_{23}+4 s_{24}\right)+26 m^4+s_{13} \left(s_{12}+s_{23}+s_{24}\right)+s_{12} \left(s_{12}+s_{14}+s_{23}+s_{24}\right)\\

0\\

-m^2 \left(11 s_{12}+6 s_{13}+4 s_{14}+11 s_{23}+4 s_{24}\right)+28 m^4+s_{13} \left(s_{12}+s_{23}+s_{24}\right)+\left(s_{12}+s_{23}\right) \left(s_{12}+s_{14}+s_{23}+s_{24}\right)\\

-m^2 \left(13 s_{12}+8 s_{13}+4 s_{14}+13 s_{23}+4 s_{24}\right)+36 m^4+\left(s_{12}+s_{23}\right) \left(s_{12}+s_{14}+s_{23}+s_{24}\right)+s_{13} \left(s_{12}+2 s_{23}+s_{24}\right)\\

m^2 \left(9 s_{12}+2 \left(2 s_{13}+s_{14}+s_{23}+2 s_{24}\right)\right)-18 m^4-s_{13} \left(s_{12}+s_{24}\right)-s_{12} \left(s_{12}+s_{14}+s_{23}+s_{24}\right)\\

m^2 \left(2 s_{12}+2 s_{13}+2 s_{14}+9 s_{23}\right)-10 m^4-s_{13} s_{23}-s_{23} \left(s_{12}+s_{14}+s_{23}+s_{24}\right)\\

-m^2 \left(11 s_{12}+6 s_{13}+2 s_{14}+4 s_{23}+4 s_{24}\right)+26 m^4+s_{13} \left(s_{12}+s_{23}+s_{24}\right)+s_{12} \left(s_{12}+s_{14}+s_{23}+s_{24}\right)\\

\end{bmatrix} \  ,\\

&e_5=\begin{bmatrix}

-2 m^2 \left(s_{12}+s_{13}+s_{23}\right)+8 m^4+s_{12} s_{13}\\

s_{12} \left(2 m^2-s_{14}\right)-\left(4 m^2-s_{23}\right) \left(2 m^2-s_{14}\right)+s_{13} \left(s_{24}-2 m^2\right)\\

-2 m^2 \left(2 s_{12}+s_{13}+s_{14}+s_{23}+s_{24}\right)+16 m^4+s_{12} s_{13}+s_{12} s_{14}\\

0\\

s_{12} \left(s_{14}-2 m^2\right)+\left(2 m^2-s_{14}\right) \left(4 m^2-s_{23}\right)+s_{13} \left(2 m^2-s_{24}\right)\\

-4 m^2 s_{23}+s_{12} \left(-4 m^2+s_{13}+s_{14}\right)+s_{14} \left(s_{23}-4 m^2\right)+16 m^4-s_{13} s_{24}\\

2 m^2 \left(-4 m^2+s_{14}+s_{24}\right)+s_{12} \left(2 m^2-s_{14}\right)\\

-2 m^2 \left(s_{13}-s_{14}-s_{23}+s_{24}\right)-s_{14} s_{23}+s_{13} s_{24}\\

-2 m^2 \left(2 s_{12}+s_{13}+s_{14}+s_{23}+s_{24}\right)+16 m^4+s_{12} s_{13}+s_{12} s_{14}\\

\end{bmatrix} \  ,
\\
&e_6=\begin{bmatrix}
-m^2 \left(7 s_{12}+2 \left(2 s_{14}+s_{23}+s_{24}\right)\right)+10 m^4+s_{12}^2+s_{14} s_{23}+s_{12} \left(s_{14}+s_{23}+s_{24}\right)\\
s_{14} \left(s_{23}-4 m^2\right)+\left(2 m^2-s_{24}\right) \left(5 m^2-s_{23}-s_{24}\right)+s_{12} \left(-2 m^2+s_{14}+s_{24}\right)\\
2 m^2 \left(5 m^2-s_{23}-s_{24}\right)+s_{12} \left(-7 m^2+s_{23}+s_{24}\right)+s_{14} \left(-4 m^2+s_{23}+s_{24}\right)+s_{12}^2\\
0\\
s_{14} \left(4 m^2-s_{23}\right)+s_{12} \left(2 m^2-s_{14}-s_{24}\right)-\left(2 m^2-s_{24}\right) \left(5 m^2-s_{23}-s_{24}\right)\\
\left(s_{12}-s_{24}\right) \left(-5 m^2+s_{12}+s_{23}+s_{24}\right)\\
s_{14} \left(s_{12}-s_{24}\right)\\
\left(2 m^2-s_{24}\right) \left(5 m^2-s_{23}-s_{24}\right)+s_{12} \left(s_{24}-2 m^2\right)+s_{14} \left(-4 m^2+s_{23}+s_{24}\right)\\
2 m^2 \left(5 m^2-s_{23}-s_{24}\right)+s_{12} \left(-7 m^2+s_{23}+s_{24}\right)+s_{14} \left(-4 m^2+s_{23}+s_{24}\right)+s_{12}^2\\
\end{bmatrix} \ .
\end{align*}
\end{tiny}
We have verified numerically that all of $e_1...e_6$ are parallel to the null vector, $e_0$, when $\epsilon(1,3,4)=0$.\\

Another choice of $q_i$ and $e_i$ is the following:
\begin{align*}
q_1&=u_1-u_2+u_3+u_5+2 u_6-u_8+u_9 \ ,\\
q_2&=u_1-u_2+u_5+2 u_6+u_7-2 u_8 \ ,\\
q_3&=2 u_1-3 u_2+2 u_3+3 u_5+5 u_6-3 u_8+2 u_9 \ ,\\
q_4&=u_1-2 u_2+2 u_3+2 u_5+3 u_6-u_7-u_8+2 u_9 \ ,\\
q_5&=-u_1+u_2-u_3-u_5-2 u_6+u_8-u_9 \ ,\\
q_6&=u_1-2 u_2+2 u_5+3 u_6+u_7-3 u_8 \ ,
\end{align*}
\begin{tiny}
 \begin{align*}
    &e_1=\begin{bmatrix}

-s_{23} \left(-13 m^2+3 s_{12}-s_{13}+s_{14}+2 s_{23}+2 s_{24}\right)\\

-2 m^2 s_{14}-7 m^2 s_{23}+s_{13} \left(8 m^2+s_{23}\right)+s_{12} \left(8 m^2-2 s_{13}+s_{23}\right)-15 m^4-s_{12}^2-s_{13}^2+s_{14}^2+2 s_{23}^2-s_{14} s_{23}+2 s_{23} s_{24}\\

-8 m^2 s_{14}+5 m^2 s_{23}-5 m^2 s_{24}+s_{12} \left(-18 m^2+2 s_{13}+2 s_{14}-s_{23}+s_{24}\right)+s_{13} \left(-8 m^2+s_{23}+s_{24}\right)+40 m^4+2 s_{12}^2-2 s_{23}^2+s_{14} s_{23}+s_{14} s_{24}-2 s_{23} s_{24}\\

-5 m^2 \left(2 s_{12}+2 s_{14}+3 s_{23}+s_{24}\right)+25 m^4+s_{12}^2-s_{13}^2+s_{14}^2+2 s_{23}^2+s_{13} s_{23}+s_{14} s_{23}+s_{13} s_{24}+s_{14} s_{24}+2 s_{23} s_{24}+s_{12} \left(2 s_{14}+3 s_{23}+s_{24}\right)\\

-m^2 \left(18 s_{12}+8 s_{13}+8 s_{14}+8 s_{23}+5 s_{24}\right)+40 m^4+2 s_{12}^2+2 s_{14} s_{23}+s_{13} s_{24}+s_{14} s_{24}+s_{12} \left(2 s_{13}+2 s_{14}+2 s_{23}+s_{24}\right)\\

-8 m^2 s_{14}+5 m^2 s_{23}-5 m^2 s_{24}+s_{12} \left(-18 m^2+2 s_{13}+2 s_{14}-s_{23}+s_{24}\right)+s_{13} \left(-8 m^2+s_{23}+s_{24}\right)+40 m^4+2 s_{12}^2-2 s_{23}^2+s_{14} s_{23}+s_{14} s_{24}-2 s_{23} s_{24}\\

-2 m^2 s_{14}-7 m^2 s_{23}+s_{13} \left(8 m^2+s_{23}\right)+s_{12} \left(8 m^2-2 s_{13}+s_{23}\right)-15 m^4-s_{12}^2-s_{13}^2+s_{14}^2+2 s_{23}^2-s_{14} s_{23}+2 s_{23} s_{24}\\

0\\

m^2 \left(-8 s_{12}-8 s_{13}+2 s_{14}+20 s_{23}\right)+15 m^4+s_{12}^2+s_{13}^2-s_{14}^2-4 s_{23}^2+2 s_{12} \left(s_{13}-2 s_{23}\right)-4 s_{23} s_{24}\\

\end{bmatrix} \  ,\\

&e_2=\begin{bmatrix}

s_{23} \left(-5 m^2+2 s_{12}+s_{14}+2 s_{24}\right)\\

m^2 \left(s_{12}+s_{13}+6 s_{14}-3 s_{23}\right)-5 m^4-s_{14}^2-s_{12} s_{14}-s_{13} s_{14}+s_{14} s_{23}+2 s_{23} s_{24}\\

m^2 \left(7 s_{12}+2 s_{13}+2 s_{14}-s_{23}+5 s_{24}\right)-10 m^4-s_{12}^2-s_{13} s_{24}-s_{14} s_{24}-s_{12} \left(s_{13}+s_{14}-s_{23}+s_{24}\right)\\

m^2 \left(8 s_{12}+3 s_{13}+8 s_{14}+s_{23}+5 s_{24}\right)-15 m^4-s_{12}^2-s_{14}^2-s_{13} s_{14}-s_{13} s_{24}-s_{14} s_{24}-s_{12} \left(s_{13}+2 s_{14}+s_{23}+s_{24}\right)\\

m^2 \left(7 s_{12}+2 s_{13}+2 s_{14}+4 s_{23}+5 s_{24}\right)-10 m^4-s_{12}^2-s_{14} s_{23}-s_{13} s_{24}-s_{14} s_{24}-2 s_{23} s_{24}-s_{12} \left(s_{13}+s_{14}+s_{23}+s_{24}\right)\\

m^2 \left(7 s_{12}+2 s_{13}+2 s_{14}-s_{23}+5 s_{24}\right)-10 m^4-s_{12}^2-s_{13} s_{24}-s_{14} s_{24}-s_{12} \left(s_{13}+s_{14}-s_{23}+s_{24}\right)\\

m^2 \left(s_{12}+s_{13}+6 s_{14}-3 s_{23}\right)-5 m^4-s_{14}^2-s_{12} s_{14}-s_{13} s_{14}+s_{14} s_{23}+2 s_{23} s_{24}\\

0\\

-m^2 \left(s_{12}+s_{13}+6 s_{14}+2 s_{23}\right)+5 m^4+s_{14}^2+s_{13} s_{14}+s_{12} \left(s_{14}+2 s_{23}\right)\\

\end{bmatrix} \  ,\\

&e_3=\begin{bmatrix}

2 m^2 s_{14}+s_{23} \left(-6 m^2+s_{12}+s_{24}\right)-2 m^4+s_{23}^2\\

-m^2 s_{24}+s_{23} \left(7 m^2-s_{12}-s_{13}-s_{24}\right)+s_{14} \left(s_{24}-2 m^2\right)+2 m^4-s_{23}^2\\

2 m^2 \left(s_{13}+s_{14}-3 \left(s_{23}+s_{24}\right)\right)-4 m^4+s_{23}^2+s_{24}^2+s_{12} s_{23}+s_{12} s_{24}+2 s_{23} s_{24}\\

s_{13} \left(2 m^2-s_{23}\right)+\left(s_{23}-s_{24}\right) \left(7 m^2-s_{12}-s_{23}-s_{24}\right)+s_{14} \left(s_{24}-2 m^2\right)\\

2 m^2 s_{13}+s_{24} \left(-6 m^2+s_{12}+s_{23}\right)-2 m^4+s_{24}^2\\

2 m^2 \left(s_{13}+s_{14}-3 \left(s_{23}+s_{24}\right)\right)-4 m^4+s_{23}^2+s_{24}^2+s_{12} s_{23}+s_{12} s_{24}+2 s_{23} s_{24}\\

-m^2 s_{24}+s_{23} \left(7 m^2-s_{12}-s_{13}-s_{24}\right)+s_{14} \left(s_{24}-2 m^2\right)+2 m^4-s_{23}^2\\

0\\

m^2 s_{24}+s_{14} \left(4 m^2-s_{24}\right)+s_{23} \left(-13 m^2+2 s_{12}+s_{13}+2 s_{24}\right)-4 m^4+2 s_{23}^2\\

\end{bmatrix} \  ,\\

&e_4=\begin{bmatrix}

s_{12} \left(-9 m^2+s_{13}+s_{14}+2 s_{23}+s_{24}\right)+\left(4 m^2-s_{23}\right) \left(5 m^2-s_{14}-s_{24}\right)+s_{13} \left(s_{24}-4 m^2\right)+s_{12}^2\\

m^2 \left(s_{12}+s_{13}+s_{14}-3 s_{23}+s_{24}\right)-5 m^4+s_{14} s_{23}-s_{14} s_{24}+s_{23} s_{24}\\

-m^2 \left(2 s_{12}+2 s_{13}+2 s_{14}+s_{23}-5 s_{24}\right)+10 m^4-s_{24}^2+s_{12} \left(s_{23}-s_{24}\right)-s_{23} s_{24}\\

m^2 \left(8 s_{12}+3 s_{13}+3 s_{14}+s_{23}+10 s_{24}\right)-15 m^4-s_{12}^2-s_{24}^2-s_{13} s_{24}-s_{14} s_{24}-s_{23} s_{24}-s_{12} \left(s_{13}+s_{14}+s_{23}+2 s_{24}\right)\\

m^2 \left(7 s_{12}+2 s_{13}+2 s_{14}+4 s_{23}+9 s_{24}\right)-10 m^4-s_{12}^2-s_{24}^2-s_{14} s_{23}-s_{13} s_{24}-2 s_{23} s_{24}-s_{12} \left(s_{13}+s_{14}+s_{23}+2 s_{24}\right)\\

-m^2 \left(2 s_{12}+2 s_{13}+2 s_{14}+s_{23}-5 s_{24}\right)+10 m^4-s_{24}^2+s_{12} \left(s_{23}-s_{24}\right)-s_{23} s_{24}\\

m^2 \left(s_{12}+s_{13}+s_{14}-3 s_{23}+s_{24}\right)-5 m^4+s_{14} s_{23}-s_{14} s_{24}+s_{23} s_{24}\\

0\\

-m^2 \left(10 s_{12}+5 s_{13}+5 s_{14}+2 s_{23}+5 s_{24}\right)+25 m^4+s_{12}^2+s_{13} s_{24}+s_{14} s_{24}+s_{12} \left(s_{13}+s_{14}+2 s_{23}+s_{24}\right)\\

\end{bmatrix} \  ,\\

&e_5=\begin{bmatrix}

s_{12} \left(-7 m^2+s_{13}+s_{14}+2 s_{23}+s_{24}\right)+\left(2 m^2-s_{23}\right) \left(5 m^2-s_{14}-2 s_{24}\right)+s_{13} \left(s_{24}-2 m^2\right)+s_{12}^2\\

-m^2 \left(s_{12}+s_{13}+s_{14}+3 s_{23}+4 s_{24}\right)+5 m^4+s_{14} s_{23}+s_{12} s_{24}+\left(s_{13}+2 s_{23}\right) s_{24}\\

\left(s_{23}-s_{24}\right) \left(s_{12}-m^2\right)\\

-\left(m^2-s_{12}\right) \left(5 m^2-s_{12}-s_{13}-s_{14}-s_{23}-s_{24}\right)\\

m^2 \left(7 s_{12}+2 s_{13}+2 s_{14}+4 s_{23}+5 s_{24}\right)-10 m^4-s_{12}^2-s_{14} s_{23}-\left(s_{13}+2 s_{23}\right) s_{24}-s_{12} \left(s_{13}+s_{14}+s_{23}+2 s_{24}\right)\\

\left(s_{23}-s_{24}\right) \left(s_{12}-m^2\right)\\

-m^2 \left(s_{12}+s_{13}+s_{14}+3 s_{23}+4 s_{24}\right)+5 m^4+s_{14} s_{23}+s_{12} s_{24}+\left(s_{13}+2 s_{23}\right) s_{24}\\

0\\

\left(m^2-s_{12}\right) \left(5 m^2-s_{12}-s_{13}-s_{14}-2 s_{23}\right)\\

\end{bmatrix} \  ,\\

&e_6=\begin{bmatrix}

-2 m^2 s_{14}+s_{23} \left(2 m^2-s_{12}-s_{24}\right)+2 m^4\\

s_{14} \left(2 m^2-s_{23}-s_{24}\right)-\left(m^2-s_{23}\right) \left(2 m^2-s_{24}\right)\\

2 m^2 \left(s_{12}+s_{23}+3 s_{24}\right)-6 m^4-s_{24}^2-s_{23} s_{24}-s_{12} \left(s_{23}+s_{24}\right)\\

s_{12} \left(2 m^2-s_{24}\right)-\left(5 m^2-s_{23}-s_{24}\right) \left(2 m^2-s_{24}\right)+s_{14} \left(4 m^2-s_{23}-s_{24}\right)\\

2 m^2 \left(s_{12}+s_{14}+3 s_{24}\right)-8 m^4-s_{24}^2-s_{12} s_{24}\\

2 m^2 \left(s_{12}+s_{23}+3 s_{24}\right)-6 m^4-s_{24}^2-s_{23} s_{24}-s_{12} \left(s_{23}+s_{24}\right)\\

s_{14} \left(2 m^2-s_{23}-s_{24}\right)-\left(m^2-s_{23}\right) \left(2 m^2-s_{24}\right)\\

0\\

-m^2 s_{24}+s_{14} \left(-4 m^2+s_{23}+s_{24}\right)+4 m^4-s_{12} s_{23}\\

\end{bmatrix} \  .\\
\end{align*}
\end{tiny}
We found numerically that for this choice all $e_i$ are parallel to $e_0$ when $\epsilon(2,3,4)=0$.

\section{Special Kinematics of 3D Topologically Massive Theories} \label{ap:Relations}
The on-shell polarisation vectors of topologically massive theories satisfy the equations of motion given in \eqref{eom ym}, while the external momenta should be on-shell, that is, $p_i^2=-m_i^2$. Furthermore, special relations can arise in three spacetime dimensions as we will see in the following. 
\begin{description}
\item[3-point Amplitudes] From  \eqref{eom ym}, we have the following relations for the polarisation vectors of the three external states 1, 2 and 3: 
\be
\begin{split}
\varepsilon_{1\mu}+\frac{i}{m}\epsilon_{\mu\nu\rho}p^{\nu}_{1}\varepsilon_{1\rho}=0\ ,\\
\varepsilon_{2\mu}+\frac{i}{m}\epsilon_{\mu\nu\rho}p^{\nu}_{2}\varepsilon_2{\rho}=0\ ,\\
\varepsilon_{3\mu}+\frac{i}{m}\epsilon_{\mu\nu\rho}p^{\nu}_{3}\varepsilon_3{\rho}=0\ .\\
\end{split}
\ee
By contracting the first line with $\varepsilon_2$ and $\varepsilon_3$ and using the second and third term respectively, we get the following relations: 
\be\label{3-point rel}
ee_{12}= -\frac{2}{m^2} ep_{21}ep_{12} \ ,\quad ee_{13}= -\frac{2}{m^2}ep_{12}ep_{32} \ .
\ee
Similarly, contracting the second line with $\varepsilon_3$ and using the last line we get:
\be
ee_{23}= -\frac{2}{m^2} ep_{21}ep_{23} \ .
\ee
This shows that
\be
ee_{ij} m^2=2 ep_{ij}ep_{ji} \ .
\ee
 We use these relations to derive \eqref{M3}.
\item[4-point Amplitudes] At 4-points, the analytic manipulations become more involve and we proceed to use a numerical approach to find the on-shell relations between polarisation vectors and momenta. We have checked that the following relations
\be\label{4-pointdc}
\begin{split}
    \frac{ee_{12}ee_{34}}{ee_{14}ee_{23}}=\frac{1}{(s+t)^4}\Bigg(&-8 m^2 s (s - 3 t) (s + t) + s^2 (s + t)^2+32 m^3 (-s + t)\sqrt{-stu}\\
    +&8 m s (s + t)\sqrt{-stu}+16 m^4 (s^2 - 6 s t + t^2)\Bigg)\ ,\\
     \frac{ee_{13}ee_{24}}{ee_{14}ee_{23}}=\frac{1}{(s+t)^4}\Bigg(&8 m^2 t (3s -  t) (s + t) + t^2 (s + t)^2+16 m^4 (s^2 - 6 s t + t^2)\\
     &-8m\sqrt{-stu}(4 m^2 (s - t) + t (s + t))\Bigg) \ ,
\end{split}
\ee
are satisfied when using random on-shell kinematics as described in Appendix \ref{num meth}. These relations can also be checked in the Breit coordiante system \eqref{breit1}, where the Mandelstam variables are related by \eqref{breit2}. Once these two relations are imposed, we can show analytically that the 4-point TMG amplitude is the double copy of the 4-point TMYM one.
\item[5-point Amplitudes] At 5-points, we will use a special property of 3D, namely, the fact that any anti-symmetric tensor with 4 indices is identically zero. It is specially useful to look at the case
\begin{equation}
    \epsilon_{[\mu\nu\rho}(p_i)_{\sigma]}=0 \ .
\end{equation}
By contracting this relation with $p_1^{\mu}p_2^{\nu}p_3^{\rho}p_4^{\sigma}$ and choosing different $i$'s, we obtain four relations between the $s_{ij}$ Mandelstam variables and $\epsilon (i,j,k)=\epsilon_{\mu\nu\sigma}p_i^{\mu}p_j^{\nu}p_k^{\sigma}$. From them, we can write $s_{12}$, $s_{13}$, $s_{14}$, $s_{23}$ in terms of $\epsilon(i,j,k)$ and $s_{24}$. For example:
\begin{small}
\begin{equation}
    s_{13}\!=\!\!\frac{\epsilon (1,2,3) \epsilon (1,3,4) \!\!\left(s_{24}\!-\!2 m^2\!\right)\!+\!m^2 \epsilon (1,2,3)^2\!\!-\!\!m^2 \!\!\left(\!\epsilon (1,2,4)^2\!-\!2 \epsilon (2,3,4) \epsilon (1,2,4)\!-\!\epsilon (1,3,4)^2\!+\!\epsilon (2,3,4)^2\!\right)}{\epsilon (1,2,4) \epsilon
   (2,3,4)} .
\end{equation}
\end{small}
To find the remaining $s_{24}$ in terms of $\epsilon(i,j,k)$ we can consider products of two $\epsilon(i,j,k)$ and expand the double $\epsilon$ in terms of the metric, that is,
\begin{equation}
\begin{split}
\epsilon (1,2,3)\epsilon (1,2,3)&=\epsilon_{\mu\nu\rho}\epsilon_{\alpha\beta\gamma}(p_{1}^{\mu}p_{2}^{\nu}p_{3}^{\rho})(p_{1}^{\alpha}p_{2}^{\beta}p_{3}^{\gamma}), \\
&=\frac{1}{4} \left(-16 m^6+8 m^4 (s_{12}+s_{13}+s_{23})-m^2 (s_{12}+s_{13}+s_{23})^2+s_{12} s_{13} s_{23}\right) \ .
\end{split}
\end{equation}
By using the previously derived relations for $s_{12}$, $s_{13}$, $s_{14}$, and $s_{23}$; we can derive an expression for $s_{24}$ purely in terms of $\epsilon(i,j,k)$.
\end{description}

\section{4-point TMG Amplitude} \label{ap:4pTMG}
In this appendix we show the explicit expression for the 4-point TMG amplitude in a general coordinate system and in the Breit coordinate system. In terms of the Mandelstam variables, the four-point amplitude of TMG reads
\begin{small}
\begin{equation*}
\begin{split}
&M_4= \Bigg(-ee_{13}^2 ee_{24}^2 \left(m^2-s\right) \Bigg(80 \left(3 m^2-s\right) \left(s-4 m^2\right)^2 m^8+4 \left(4 m^2-s\right) \left(1384 m^6-897 s m^4+179 s^2 m^2-8 s^3\right) t m^4\\
&+\left(22128 m^6-5568 s m^4+729 s^2 m^2-119 s^3\right) t^3 m^2-4 \left(7552 m^8-3258 s m^6+568 s^2 m^4-85 s^3 m^2+8 s^4\right) t^2 m^2\\
&-6 \left(4 m^2+s\right) t^6+\left(2 m^2-s\right) \left(76 m^2+3 s\right) t^5-\left(5360 m^6-916 s m^4+154 s^2 m^2+s^3\right) t^4\Bigg) s^2\\
&+ee_{12} ee_{34} \left(m^2-t\right) t \Bigg(ee_{14} ee_{23} \Bigg(-1477632 m^{16}+256 (5182 s+6493 t) m^{14}-64 \left(5444 s^2+15491 t s+10889 t^2\right) m^{12}\\
&+16 \left(984 s^3+3973 t s^2+12776 t^2 s+7851 t^3\right) m^{10}+4 \left(5760 s^4+10402 t s^3+13539 t^2 s^2+1045 t^3 s-1568 t^4\right) m^8\\
&-4 \left(1600 s^5+5084 t s^4+5866 t^2 s^3+3305 t^3 s^2+971 t^4 s+151 t^5\right) m^6\\
&+\left(512 s^6+1976 t s^5+4732 t^2 s^4+2798 t^3 s^3+152 t^4 s^2-11 t^5 s+24 t^6\right) m^4+\\
&t \left(128 s^6+360 t s^5+232 t^2 s^4-249 t^3 s^3+26 t^4 s^2+22 t^5 s+4 t^6\right) m^2+s t^3 \left(-2 s^4-5 t s^3+10 t^2 s^2+2 t^3 s+t^4\right)\Bigg)\\
&+ee_{12} ee_{34} \Bigg(-768 (486 s-481 t) m^{14}+64 \left(3360 s^2-49 t s-6493 t^2\right) m^{12}\\
&+16 \left(-600 s^3-613 t s^2+10912 t^2 s+10889 t^3\right) m^{10}-4 \Bigg(3072 s^4+7778 t s^3+8855 t^2 s^2+15613 t^3 s\\
&+7851 t^4\Bigg) m^8+4 \left(1088 s^5+1948 t s^4+2314 t^2 s^3+491 t^3 s^2+1282 t^4 s+392 t^5\right) m^6\\
&+\left(-512 s^6-952 t s^5-492 t^2 s^4+2178 t^3 s^3+1192 t^4 s^2+434 t^5 s+151 t^6\right) m^4\\
\end{split}
\end{equation*}
\be\label{M4}
\begin{split}
&-t \left(128 s^6+488 t s^5+696 t^2 s^4+343 t^3 s^3+27 t^4 s^2+29 t^5 s+6 t^6\right) m^2\\
&-t^3 (2 s+t) \left(-s^4-4 t s^3+t^2 s^2+t^3 s+t^4\right)\Bigg)\Bigg)\\
&+ee_{13} ee_{24} \Bigg(ee_{14} ee_{23} \left(s-m^2\right) \Bigg(-8 \left(s-4 m^2\right)^2 \left(s-3 m^2\right) \left(3 s-22 m^2\right) m^6\\
&-4 \left(s-4 m^2\right) \left(-3064 m^6+2221 s m^4-517 s^2 m^2+36 s^3\right) t m^4+2 \Bigg(29184 m^8-17444 s m^6+3976 s^2 m^4-437 s^3 m^2\\
&+21 s^4\Bigg) t^2 m^2+6 \left(4 m^2+s\right) t^6+\left(-248 m^4+118 s m^2-3 s^2\right) t^5+\left(5392 m^6-1028 s m^4+204 s^2 m^2-5 s^3\right) t^4\\
&+\left(-30256 m^8+10336 s m^6-1541 s^2 m^4+163 s^3 m^2-2 s^4\right) t^3\Bigg) s^2+ee_{12} ee_{34} \Bigg(768 \left(5 s^2+486 t s-481 t^2\right) m^{16}\\
&-64 \left(92 s^3+4229 t s^2+8651 t^2 s-10822 t^3\right) m^{14}+16 \left(123 s^4+1027 t s^3+29747 t^2 s^2+14110 t^3 s-26039 t^4\right) m^{12}\\
&+4\left(66 s^5+7447 t s^4-11718 t^2 s^3-62062 t^3 s^2-15385 t^4 s+25368 t^5\right) m^{10}\\
&-4 \left(52 s^6+2394 t s^5+4948 t^2 s^4+2145 t^3 s^3-16136 t^4 s^2-4706 t^5 s+1901 t^6\right) m^8\\
&+\left(24 s^7+716 t s^6+6422 t^2 s^5+13743 t^3 s^4+9891 t^4 s^3-7861 t^5 s^2-2569 t^6 s-484 t^7\right) m^6\\
&+t \left(16 s^7-972 t s^6-2312 t^2 s^5-2829 t^3 s^4-25 t^4 s^3-317 t^5 s^2-73 t^6 s+36 t^7\right) m^4\\
&+t^2 \left(86 s^7+263 t s^6+259 t^2 s^5-50 t^3 s^4-100 t^4 s^3+23 t^5 s^2+13 t^6 s+4 t^7\right) m^2\\
&+s t^3 \left(2 s^6+9 t s^5+15 t^2 s^4+20 t^3 s^3+6 t^4 s^2+3 t^5 s+t^6\right)\Bigg)\Bigg)\Bigg)\times\\
&\frac{-i}{128 m^2 (m^2 - s) s^2 (m^2 - t) t (-4 m^2 + s + t)^2 (-3 m^2 + s + t)} \ .
\end{split}
\ee
\end{small}
In order to see the double copy relation explicitly, we write this amplitude in the Breit coordinate system which was defined in \eqref{breit1}. The amplitude largely simplifies and is now given by
\begin{small}
\be\label{4pt Breit}
\begin{split}
M_4=&\frac{-m}{2 \lambda ^2 p^2 \left(\lambda ^2+m^2\right)^2 \left(m^2+4 p^2\right) \left(3 m^4+4 m^2 \left(\lambda ^2+p^2\right)+4 \lambda ^2 p^2\right)}\times\\
\Bigg(&30 E \lambda  m^{10} \left(\lambda ^2 p-4 p^3\right)+3 i m^{11} \left(\lambda ^4+16 p^4-96 \lambda ^2 p^2\right)+8 E \lambda ^5 p^5 \left(\lambda ^4-32 p^4-4 \lambda ^2 p^2\right)\\
&+E m^8 \left(-880 \lambda  p^5+366 \lambda ^3 p^3+88 \lambda ^5 p\right)+8 E \lambda ^3 m^2 p^3 \left(-8 \lambda ^6+320 p^6+8 \lambda ^2 p^4+\lambda ^4 p^2\right)\\
&+2 i m^9 \left(5 \lambda ^6+152 p^6-502 \lambda ^2 p^4-367 \lambda ^4 p^2\right)+2 i \lambda ^4 m p^4 \left(18 \lambda ^6+640 p^6+416 \lambda ^2 p^4-17 \lambda ^4 p^2\right)\\
& +2 E \lambda  m^6 p \left(41 \lambda ^6-960 p^6+856 \lambda ^2 p^4+148 \lambda ^4 p^2\right)+i m^7 \left(11 \lambda ^8+512 p^8-3396 \lambda ^2 p^6-667 \lambda ^4 p^4-652 \lambda ^6 p^2\right)\\
&+2 E \lambda  m^4 p \left(12 \lambda ^8-640 p^8+1776 \lambda ^2 p^6+552 \lambda ^4 p^4-131 \lambda ^6 p^2\right)\\
&-i \lambda ^2 m^3 p^2 \left(56 \lambda ^8+2560 p^8-768 \lambda ^2 p^6-1352 \lambda ^4 p^4-77 \lambda ^6 p^2\right)\\
&+2 i m^5 \left(2 \lambda ^{10}+128 p^{10}-2464 \lambda ^2 p^8-413 \lambda ^4 p^6+217 \lambda ^6 p^4-131 \lambda ^8 p^2\right)\Bigg) \ .
\end{split}
\ee
\end{small}
In this coordinate system, the shifted kinematic factors of TMYM, \eqref{shift n}, read
\begin{small}
\begin{equation}\label{breit n}
    \begin{split}
        \hat{n}_s=&\frac{-i\lambda }{m p \left(p^2-E^2\right) \left(-E^2+m^2+p^2\right)}\times\\
        \Bigg(&-5 \lambda  m^5 p-\lambda  m^3 p \left(\lambda ^2+31 p^2\right)-2 i \lambda ^2 p^2 \sqrt{m^2+p^2} \left(\lambda ^2+4 p^2\right)+i m^4 \sqrt{m^2+p^2} \left(\lambda ^2+2 p^2\right)\\
        &m \left(-16 \lambda  p^5-9 \lambda ^3 p^3+4 \lambda ^5 p\right)++i m^2 \sqrt{m^2+p^2} \left(\lambda ^4+8 p^4+2 \lambda ^2 p^2\right)\\
        &E\Bigg(2 i m^4 \left(\lambda ^2+5 p^2\right)+\lambda  m p \sqrt{m^2+p^2} \left(3 \lambda ^2-16 p^2\right)+5 \lambda  m^3 p \sqrt{m^2+p^2}-2 i \lambda ^2 p^2 \left(\lambda ^2+4 p^2\right)\\
        &+2 i m^2 \left(\lambda ^4+4 p^4-7 \lambda ^2 p^2\right)\Bigg)\Bigg) \ ,\\
         \hat{n}_t=&-\frac{2  \lambda  \sqrt{m^2+p^2}}{m p \left(p^2-E^2\right) \left(-E^2+m^2+p^2\right)}\times\\
         \Bigg(&-5 i E \lambda  m^3 p+i E \lambda  m \left(16 p^3-3 \lambda ^2 p\right)+m^4 \left(\lambda ^2+2 p^2\right)+m^2 \left(\lambda ^4+8 p^4+2 \lambda ^2 p^2\right)-2 \lambda ^2 p^2 \left(\lambda ^2+4 p^2\right)\Bigg)\ ,\\
         \hat{n}_{u}=&-\hat{n}_s-\hat{n}_t \ .
    \end{split}
\end{equation}
\end{small}
As mentioned in the bulk of the paper, by plugging in these kinematic factors in \eqref{4-point dc} and using \eqref{breit2}, we get the amplitude of TMG \eqref{4pt Breit}.

\section{Numerical Method}\label{num meth}
In this section we explain how we generate the numerical 3D on-shell kinematics and give some specific values that we used to check the double copy of TMYM at 5-point. We perform all computations over finite fields, which have a finite number of elements. This way we avoid numerical errors and make the calculations more efficient. We consider the field of integers modulo p, a prime number which is equal to $p=2147483497$ in this paper. Computations over finite fields are common and have been used to reconstruct polynomials in kinematics variables in loop QCD calculations \cite{Peraro:2016wsq,Zeng:2017ipr,Badger:2017jhb,Abreu:2019odu}. We refer the reader to \cite{ Laurentis:2019bjh, Peraro:2016wsq} for a detailed explanation.\\

We generate random kinematics such that the two following conditions are satisfied:

\begin{itemize}
    \item The momenta are on-shell and conserved.
    \item The polarisation vectors satisfy the constraint \eqref{eom ym}.
\end{itemize}
The components of each external momenta are related by the on-shell condition,
\begin{equation}
    p^{\mu}=(p_0,p_1,p_2)\ , \quad -p^2_{0}+p^2_{1}+p^2_{2}=-m^2\ ,
\end{equation}
and the second condition is satisfied for the polarisation vector built out of these components as,
\be\label{pola values}
\epsilon^{\mu}=\Bigg(-\frac{(m-p_{0}-p_{1}+ip_{2})(m+p_{0}+p_{1}+ip_{2})}{2m(p_{0}+p_{1})},\frac{(p_{0}+p_{1})^2+(m+ip_{2})^2}{2m(p_{0}+p_{1})},-i+\frac{p_2}{m}\Bigg)\ .
\ee
We have included ten of the random kinematics that we used to calculate the 5-point TMG in Table \ref{tab:valuesMandelstam} and the unshifted numerators of TMYM in Table 2. The unshifted numerators of TMYM double copy to the 5-point TMG amplitude using \eqref{eq:Mn final}.

\begin{table}[H]
    \centering
    \begin{tabular}{|c|c|c|c|c|c||c|}
    \hline
         $ $& $m$ & $s_{12} $ & $s_{13} $ & $s_{14} $ & $s_{23} $& $M^{TMG}_{5}$ \\
         \hline\hline

         1 & $384817470$ & $1158823430$ & $345329619$ & $1397768610$ & $264965055$ & $1010590219\kappa^3$\\
         \hline
        2  & $1250040736$& $652270246$ & $1821369346$ & $1622372086$ & $739244825$ & $1355550730 \kappa^3$\\
        \hline
        3 & $800857604$ & $2035880423$ & $1968515133$ & $1224440350$ & $664321872$ & $526697979 \kappa^3$\\
        \hline
        4  & $1150467713$ & $2060321774$ & $82539557$ & $702220445$ & $431821399$ & $467871508\kappa^3$\\
        \hline
        5  & $158667339$& $2051154971$ & $369848949$ & $890093650$ & $756917203$ & $475230586 \kappa^3$\\
        \hline
        6  & $1916307032$& $541901353$ & $2099692150$ & $150737937$ & $425995603$ & $1278139921 \kappa^3$\\        
        \hline    
        7  & $1662283157$& $938971574$ & $50758705$ & $928659888$ & $1820858158$ & $ 108642017\kappa^3$\\        
        \hline    
        8  & $1078072319$ & $1151859367$ & $1186765675$ & $110159710$ & $209051438$ & $ 1638775080\kappa^3$\\        
        \hline  
        9  & $231108131$ & $1439516500$ & $572657143$ & $405624245$ & $68286568$ & $266492730 \kappa^3$\\        
        \hline  
        10  & $1849710816$ & $247156271$ & $155877255$ & $2085263836$ & $62583717$ & $966750502 \kappa^3$\\        
        \hline  
    \end{tabular}
    \caption{Examples of the kinematic values used to calculate the unshifted numerators of TMYM and the 5-point TMG. The values are in the field of integers modulo $p=2147483497$. The remaining Mandelstam variable, $s_{24}$, can be obtained by requiring that \eqref{eq:det99} is zero in 3D.}
    \label{tab:valuesMandelstam}
\end{table}

\begin{table}[H]
    \centering
    \resizebox{\textwidth}{!}{%
    \begin{tabular}{|c|c|c|c|c|c|c|c|c|c|c|}
    \hline
         $$ & $1$ & $2 $ & $3 $ & $4 $ & $5 $& $6$ & $7$& $8$& $9$& $10$\\
         \hline\hline

         $n_{1}$ & $1679102633$ & $348983868$ & $399241281$ & $842732794$ & $300495714$ & $332245457$ & $2078791969$ & $405757021$ & $704097804$& $1147807466$\\
         \hline
          $n_{2}$ & $317552067$ & $1079962755$ & $683351824$ & $2113803420$ & $650623635$ & $1651377807$ & $443295016$ & $1307896649$ & $546715501$ & $33350122$\\
         \hline
          $n_{3}$ & $1771495024$ & $671539061$ & $121827398$ & $1212918710$ & $877929848$ & $1648524257$ & $1751453994$ & $715853795$ & $437788195$& $1673049330$\\
         \hline
          $n_{4}$& $1358547387$ & $1615179844$ & $1720845160$ & $1032631735$ & $698981299$ & $676408057$ & $1421711641$ & $586269609$ & $1143032868$& $1144909226$\\
         \hline
          $n_{5}$& $211933474$ & $81190884$ & $921568849$ & $1552408865$ & $1716179733$ & $1587978966$ & $1832576367$ & $503370947$ & $1216904523$& $1157374746$\\
         \hline
          $n_{6}$& $1117522969$ & $1640095150$ & $1919642084$ & $998205023$ & $1533934317$ & $1763892991$ & $1547349116$ & $1008373879$ & $1419527987$& $1656418171$\\
         \hline
          $n_{7}$ & $1479358736$ & $259089853$ & $155577739$ & $1231349959$ & $1543395860$ & $88931139$ & $114315550$ & $998844655$ & $1430903268$& $310456086$\\
         \hline
          $n_{8}$& $1654319285$ & $823920651$ & $859957704$ & $2079622430$ & $322878784$ & $1433806805$ & $651369019$ & $638654072$ & $1572317114$& $1484455868$\\
         \hline
          $n_{9}$ & $925637889$ & $1084982689$ & $1510593127$ & $1223632823$ & $991858505$ & $1585631481$ & $1130701506$ & $1854538988$ & $845822321$& $107084633$\\
         \hline
          $n_{10}$& $241485408$ & $181500425$ & $2095141802$ & $1445698561$ & $946176777$ & $1036495597$ & $924128220$ & $665606563$ & $1436242466$& $167117366$\\
         \hline
          $n_{11}$& $1913208290$ & $547913677$ & $1461674584$ & $601184017$ & $391951191$ & $112316195$ & $407293624$ & $777544715$ & $908354210$& $719884700$\\
         \hline
          $n_{12}$& $1372613275$ & $659324877$ & $523294703$ & $785456003$ & $1120871918$ & $2024095541$ & $618177398$ & $2047361914$ & $1454336818$& $1233365105$\\
         \hline
          $n_{13}$ & $1789815668$ & $1037075753$ & $209962576$ & $493974229$ & $1279666813$ & $1941726328$ & $1377397281$ & $1546104552$ & $2031446261$& $1379695040$\\
         \hline
          $n_{14}$& $1244173774$ & $1475461191$ & $1520770510$ & $1520373298$ & $1781914190$ & $211939090$ & $1680738019$ & $1229996023$ & $591423398$& $2134631466$\\
         \hline
          $n_{15}$ & $1806462521$ & $318300183$ & $733389430$ & $1399767053$ & $630210267$ & $1986264353$ & $1729317645$ & $1452460649$ & $488881236$& $1508026444$\\
         \hline
    \end{tabular}}
     \caption{Numerical values for the unshifted kinematic factors of the 5-point TMYM.}
    \label{tab: numerator values}
\end{table}

\section{BCJ Relation in Terms of Partial Amplitudes}
The BCJ relation in terms of colour ordered partial amplitudes takes the following form:
\be \label{eq:BCJpartialA}
U.
\begin{bmatrix} 
A_5[12345]\\
A_5[12435]\\
A_5[13245]\\
A_5[13425]\\
A_5[14235]\\
A_5[14325]
\end{bmatrix}=0 \ ,
\ee
where $U=\{u_1,...,u_6\}$ in this basis is given as:
\begin{align*}
u_1&=\left(m^2-s_{12}\right) \Big(\epsilon (1,2,4) \left(-5 m^2+s_{12}+s_{14}+s_{24}\right)+m^2 \epsilon (1,2,3)-s_{13} \epsilon (1,3,4)+s_{24} \epsilon (1,3,4)\\&-s_{14} \epsilon (2,3,4)+s_{23} \epsilon (2,3,4)\Big) \ ,
\end{align*}
\begin{align*}
u_2&=7 m^2 s_{24} \epsilon (1,2,3)+m^2 s_{14} \epsilon (1,2,4)+m^2 s_{24} \epsilon (1,2,4)+3 m^2 s_{14} \epsilon (1,3,4)+2 m^2 s_{24} \epsilon (1,3,4)\\&-m^2 s_{24} \epsilon (2,3,4)+s_{23} \left(m^2 (\epsilon (1,2,3)+3 \epsilon (1,3,4)-\epsilon (2,3,4))-s_{14} \epsilon (1,3,4)-s_{24} (\epsilon (1,2,3)+\epsilon (1,3,4))\right)\\&+s_{12} \Big(3 m^2 \epsilon (1,2,3)+4 m^2 \epsilon (1,2,4)+4 m^2 \epsilon (1,3,4)-9 m^2 \epsilon (2,3,4)-s_{23} \epsilon (1,3,4)-s_{14} (\epsilon (1,2,4)+\epsilon (1,3,4)\\&-\epsilon (2,3,4))-s_{24} (2 \epsilon (1,2,3)+\epsilon (1,2,4)+\epsilon (1,3,4)-\epsilon (2,3,4))+s_{23} \epsilon (2,3,4)\Big)-7 m^4 \epsilon (1,2,3)\\&-3 m^4 \epsilon (1,2,4)-5 m^4 \epsilon (1,3,4)+7 m^4 \epsilon (2,3,4)-s_{24}^2 \epsilon (1,2,3)-s_{14} s_{24} \epsilon (1,3,4)-s_{12}^2 (\epsilon (1,2,4)+\epsilon (1,3,4)
\\&-2 \epsilon (2,3,4))-s_{14} s_{24} \epsilon (2,3,4) \ ,
\end{align*}
\begin{align*}
u_3&=7 m^2 s_{24} \epsilon (1,2,3)+m^2 s_{14} \epsilon (1,2,4)-3 m^2 s_{13} \epsilon (1,3,4)-4 m^2 s_{24} \epsilon (1,3,4)-m^2 s_{13} \epsilon (2,3,4)-m^2 s_{14} \epsilon (2,3,4)\\&-m^2 s_{24} \epsilon (2,3,4)+s_{12} \Big(2 m^2 \epsilon (1,2,3)+5 m^2 \epsilon (1,2,4)-3 m^2 \epsilon (1,3,4)-5 m^2 \epsilon (2,3,4)-s_{24} (2 \epsilon (1,2,3)+\epsilon (1,2,4))\\&+s_{13} \epsilon (1,3,4)+s_{14} (\epsilon (2,3,4)-\epsilon (1,2,4))\Big)+s_{23} \Big(m^2 (2 \epsilon (1,2,3)-3 \epsilon (1,3,4)-\epsilon (2,3,4))+s_{24} \big(\epsilon (1,3,4)\\&-\epsilon (1,2,3)\big)+s_{13} (\epsilon (1,3,4)+\epsilon (2,3,4))\Big)-8 m^4 \epsilon (1,2,3)-2 m^4 \epsilon (1,2,4)+11 m^4 \epsilon (1,3,4)+7 m^4 \epsilon (2,3,4)\\&-s_{24}^2 \epsilon (1,2,3)+s_{13} s_{24} \epsilon (1,3,4)+s_{12}^2 (\epsilon (2,3,4)-\epsilon (1,2,4)) \ ,
\end{align*}
\begin{align*}
u_4=(\epsilon (1,2,4)+\epsilon (1,3,4)-\epsilon (2,3,4)) \left(-\left(m^2-s_{12}\right)\right) \left(4 m^2-s_{12}-s_{14}-s_{24}\right)\ ,
\end{align*}
\begin{align*}
u_5&=m^2 s_{14} (\epsilon (1,2,4)+2 \epsilon (1,3,4))+s_{12} \Big(m^2 (\epsilon (1,2,3)+5 \epsilon (1,2,4)+5 \epsilon (1,3,4)-4 \epsilon (2,3,4))\\&-s_{24} (\epsilon (1,2,3)+\epsilon (1,2,4)+\epsilon (1,3,4))-s_{14} (\epsilon (1,2,4)+\epsilon (1,3,4)-\epsilon (2,3,4))\Big)+s_{24} \Big(m^2 (2 \epsilon (1,2,3)+2 \epsilon (1,2,4)\\&+3 \epsilon (1,3,4)+\epsilon (2,3,4))-s_{14} (\epsilon (1,3,4)+\epsilon (2,3,4))\Big)+m^4 (-(2 \epsilon (1,2,3)+5 \epsilon (1,2,4)+6 \epsilon (1,3,4)-2 \epsilon (2,3,4)))\\&-s_{12}^2 (\epsilon (1,2,4)+\epsilon (1,3,4)-\epsilon (2,3,4))\ ,
\end{align*}
\begin{align*}
u_6&=
m^2 \Big(-4 m^2 \epsilon (1,2,3)-3 m^2 \epsilon (1,2,4)+6 m^2 \epsilon (1,3,4)+6 m^2 \epsilon (2,3,4)-2 s_{13} \epsilon (1,3,4)+s_{24} (2 \epsilon (1,2,3)+\epsilon (1,2,4)\\&+\epsilon (1,3,4))+s_{14} (\epsilon (1,2,4)-\epsilon (2,3,4))-s_{13} \epsilon (2,3,4)\Big)+s_{12} \Big(m^2 \epsilon (1,2,3)+5 m^2 \epsilon (1,2,4)-2 m^2 \epsilon (1,3,4)\\&-5 m^2 \epsilon (2,3,4)+s_{13} \epsilon (1,3,4)-s_{24} (\epsilon (1,2,3)+\epsilon (1,2,4)+\epsilon (1,3,4))+s_{14} (\epsilon (2,3,4)-\epsilon (1,2,4))\Big)\\&+s_{23} \left(m^2 (\epsilon (1,2,3)-2 \epsilon (1,3,4)-\epsilon (2,3,4))+s_{13} (\epsilon (1,3,4)+\epsilon (2,3,4))\right)+s_{12}^2 (\epsilon (2,3,4)-\epsilon (1,2,4)) \ .
\end{align*}

\bibliographystyle{JHEP}
\bibliography{references}

\end{document}